%% file: main.tex
\documentclass[sigconf, noacm]{acmart}

\AtBeginDocument{%
  }

\renewcommand\footnotetextcopyrightpermission[1]{}
\setcopyright{acmlicensed}
\copyrightyear{2025}
\acmYear{2025}
\acmConference[VLDB'25]{51st International Conference on Very Large Data Bases}{September 1--5, 2025}{London, United Kingdom}

\usepackage{tabularx}
\usepackage{algorithm}
\usepackage{algpseudocodex}
\usepackage[clock]{ifsym}

\newcommand{\sparagraph}[1]{\vspace{1mm}\noindent {\bf #1}}

\newcommand{\Systime}{System runtime}
\newcommand{\systime}{\MakeLowercase{\Systime{}}}
\newcommand{\Fulltime}{End-to-end runtime}
\newcommand{\fulltime}{\MakeLowercase{\Fulltime{}}}
\newcommand{\Qvec}{Query feature vector}
\newcommand{\qvec}{\MakeLowercase{\Qvec{}}}
\newcommand{\TSvec}{Interaction feature vector}
\newcommand{\tsvec}{\MakeLowercase{\TSvec{}}}

\newcommand\vldbdoi{XX.XX/XXX.XX}

\newcommand{\sysname}{\textit{IconqSched}}
\newcommand{\modelname}{\textit{Iconq}}

\begin{document}
\title{Improving DBMS Scheduling Decisions with Fine-grained Performance Prediction on Concurrent Queries -- Extended}

%%
%% The "author" command and its associated commands are used to define the authors and their affiliations.
\settopmatter{authorsperrow=4}
\author{Ziniu Wu}
\affiliation{
  \institution{MIT CSAIL}
  \city{Cambridge}
  \state{MA, USA}
}
\email{ziniuw@mit.edu}

\author{Markos Markakis}
\affiliation{
  \institution{MIT CSAIL}
  \city{Cambridge}
  \state{MA, USA}
}
\email{markakis@mit.edu}

\author{Chunwei Liu}
\affiliation{
  \institution{MIT CSAIL}
  \city{Cambridge}
  \state{MA, USA}
}
\email{chunwei@csail.mit.edu}

\author{Peter Baile Chen}
\affiliation{
  \institution{MIT CSAIL}
  \city{Cambridge}
  \state{MA, USA}
}
\email{peterbc@mit.edu}

\author{Balakrishnan (Murali) Narayanaswamy}
\affiliation{
  \institution{Amazon Web Services}
  \city{Santa Clara}
  \state{CA, USA}
}
\email{muralibn@amazon.com}

\author{Tim Kraska}
\affiliation{
  \institution{Amazon Web Services, \break MIT CSAIL}
  \city{Cambridge}
  \state{MA, USA}
}
\email{kraska@mit.edu}

\author{Samuel Madden}
\affiliation{
  \institution{MIT CSAIL}
  \city{Cambridge}
  \state{MA, USA}
}
\email{madden@csail.mit.edu}

\renewcommand{\shortauthors}{Ziniu Wu et al.}

\begin{abstract}

Query scheduling is a critical task that directly impacts query performance in database management systems (DBMS). Deeply integrated schedulers, which require changes to DBMS internals, 
are usually customized for a specific engine and can take months to implement. In contrast, non-intrusive schedulers make coarse-grained decisions, such as controlling query admission and re-ordering query execution, without requiring modifications to DBMS internals. They require much less engineering effort and can be applied across a wide range of DBMS engines, offering immediate benefits to end users. However, most existing non-intrusive scheduling systems rely on simplified cost models and heuristics that cannot accurately model query interactions under concurrency and different system states, possibly leading to suboptimal scheduling decisions.

This work introduces \sysname{}, a new, principled non-intrusive scheduler that optimizes the execution order and timing of queries to enhance total \fulltime{} as experienced by the user --- query queuing time plus \systime{}. Unlike previous approaches, \sysname{} features a novel fine-grained predictor, \modelname{}, which treats the DBMS as a black box and accurately estimates the \systime{} of concurrently executed queries under different system states. Using these predictions, \sysname{} is able to capture \systime{} variations across different query mixes and system loads. It then employs a greedy scheduling algorithm to effectively determine which queries to submit and when to submit them. We compare \sysname{} to other schedulers in terms of \fulltime{} using real workload traces. On Postgres, \sysname{} reduces \fulltime{} by 16.2\%-28.2\% on average and 33.6\%-38.9\% in the tail. Similarly, on Redshift, it reduces \fulltime{} by 10.3\%-14.1\% on average and 14.9\%-22.2\% in the tail.

\end{abstract}

\maketitle

%\pagestyle{\vldbpagestyle}
%\begingroup\small\noindent\raggedright\textbf{PVLDB Reference Format:}\\
%\vldbauthors. \vldbtitle. PVLDB, \vldbvolume(\vldbissue): \vldbpages, \vldbyear.\\
%\href{https://doi.org/\vldbdoi}{doi:\vldbdoi}
%\endgroup
%\begingroup
%\renewcommand\thefootnote{}\footnote{\noindent
%This work is licensed under the Creative Commons BY-NC-ND 4.0 International License. Visit \url{https://creativecommons.org/licenses/by-nc-nd/4.0/} to view a copy of this license. For any use beyond those covered by this license, obtain permission by emailing \href{mailto:info@vldb.org}{info@vldb.org}. Copyright is held by the owner/author(s). Publication rights licensed to the VLDB Endowment. \\
%\raggedright Proceedings of the VLDB Endowment, Vol. \vldbvolume, No. \vldbissue\ %
%ISSN 2150-8097. \\
%\href{https://doi.org/\vldbdoi}{doi:\vldbdoi} \\
%}\addtocounter{footnote}{-1}\endgroup


%\ifdefempty{\vldbavailabilityurl}{}{
%\vspace{.3cm}
%\begingroup\small\noindent\raggedright\textbf{PVLDB Artifact Availability:}\\
%The source code, data, and/or other artifacts have been made available at \url{\vldbavailabilityurl}.
%\endgroup

\input{intro}

\input{system}

\input{cost_mode}

\input{scheduling}

\input{experiment}

\input{background}

\input{conclusion}

\clearpage

\bibliographystyle{ACM-Reference-Format}
\bibliography{main}

\clearpage
\appendix
\input{appendix}

\end{document}

%% file: intro.tex
\section{Introduction}
\label{sec:intro}

%\markos{General comment: Do we want to name this approach? It will be better for the eval (instead of calling it just "ours") and also for people comparing against us in the future}
%\ziniuw{Yes, currently set a random name suggested by ChatGPT \sysname{}. We should figure out better names.}

%\noindent \textbf{Background}

Query scheduling is a fundamental problem in database management systems (DBMSs) that directly impacts query performance.
A scheduler needs to decide when to execute a query (or operators of a query) and how much of each resource (e.g., CPU, memory) to allocate. 
Query scheduling is challenging because making an optimal decision requires accurately predicting the cost/runtime of executing different operators with access to varying amounts of resources and under different system loads and concurrency levels~\cite{sabek2022lsched}.  Even with accurate predictions, scheduling itself is an NP-complete decision problem~\cite{ullman1975np}. 
%\srm{fix the following -- i don't know that it's fair to say it's more important just that we focus on it in this paper.} Good scheduling decisions are most crucial when handling OLAP workloads since their impact can be more significant for long-running OLAP queries than short-running OLTP queries.

This work focuses on scheduling OLAP queries because they involve accessing large amounts of data and resources. Thus, optimally scheduling each OLAP query individually can lead to more performance gain than OLTP query. 
%\srm{Why only OLAP?} \markos{the main reason is that we need long enough runtimes to have query interaction and scheduling actually make a difference}
Several past efforts have built deeply integrated DBMS schedulers~\cite{kelley1959critical, peha1996cost, chi2011icbs, grandl2014multi, grandl2016graphene, sabek2022lsched, mao2019learning, lyu2022fine, SaxenaRCLCCMKPN23} for OLAP queries, which require modifications to the DBMS internals, such as changing the way the system allocates resources at the query operator level.
This requirement limits applicability, because the schedulers are tailored to a specific DBMS, making generalization to other engines non-trivial.
In addition, it can take months for engineers to implement and verify the performance of such schedulers before they can be used by available to  end users. 
To address these shortcomings, \textit{non-intrusive} schedulers~\cite{denning1980working, zhang2012discovering, haritsa1993value, chi2013distribution, peha1991scheduling, ahmad2011interaction} make coarse-grained decisions, such as controlling query admission and reordering query execution, and as such, do not require modifications to DBMS internals. They require much less engineering effort and can potentially have an immediate impact on the end users of almost any DBMS engine~\cite{zhang2017workload, mehta2008bi}. However, most of the existing solutions (e.g., first-in-first-out, shortest-query-first~\cite{SaxenaRCLCCMKPN23}, fair scheduling~\cite{gupta2009fair, ghodsi2011dominant}) are based on simplified cost/runtime models and heuristics that can easily lead to inaccurate predictions and suboptimal scheduling decisions.

%\smallskip
In this work, we propose \sysname{}, an improved non-intrusive scheduler made possible by \modelname{}, a new fine-grained prediction model that is able to accurately predict the \emph{\systime{}} of concurrent queries (i.e. the time they take from submission to the DBMS until completion). As a result, \sysname{} can accurately understand query performance under different concurrent query loads and levels of resource availability.  Using this model, \sysname{} then schedules the submission time of each query on the host DBMS to maximally improve overall \emph{\fulltime{}} over the workload, including both \systime{} and query queuing time. Building \sysname{} poses two technical challenges:

\textbf{Challenge 1: Accurately predicting the \systime{}s of concurrently executing queries.} Building an accurate concurrent query \systime{} predictor is challenging because it needs to understand complex query interactions and account for different system states (e.g., resource utilization levels and cache state), which significantly affect query performance. Therefore, most existing approaches use simplifying assumptions or heuristics that can lead to substantial estimation errors~\cite{wu2013towards, duggan2014contender, ahmad2011predicting, ahmad2009query, duggan2011performance, ahmad2011interaction}. 
%\srm{Why are we calling out this paper specifically - it doesn't sound like it works all that great?} \ziniuw{Because it is a learned method so it does not really fall into the previous category. Also people may know that work and wonder what is the difference} 
A recent learned method~\cite{zhou2020query} uses a graph neural network (GNN) to featurize the physical query plan and focus on modeling the data sharing and memory contention among queries. However, it does not model the system state changes nor the impact of the query submission time, often producing inaccurate estimations. Furthermore, it assumes perfect knowledge of future queries to model the query interaction, which is impractical in our settings.

\textbf{Challenge 2: Deciding when and which query to execute, based on predictions.} Even given accurate predictions, \sysname\ still needs to make a non-trivial decision about which queries to execute and when. A brute-force approach will enumerate the entire search space of possible decisions, including an exponential number of combinations of candidate queries to submit concurrently and infinitely many candidate timestamps to submit them at. Thus, building a more efficient approach is necessary.

\sparagraph{Our Approach.} We develop \sysname{} as a lightweight proxy layer that can be implemented outside a DBMS instance to help schedule users' queries effectively. In principle, \sysname{} can be applied to any relational DBMS.

To address \textbf{Challenge 1}, we develop a novel \systime{} predictor, \modelname{} (\underline{I}nteraction of \underline{con}current \underline{q}ueries), which accurately models complex query interactions and system state changes. \modelname{} first encodes a stream or batch of queries as \tsvec{}s, which can capture various complex query interactions.
Then, it uses a bi-directional LSTM model to ingest these \tsvec{}s and predict the \systime{} of concurrently executed queries. 
The LSTM model's internal states implicitly encode the concurrent state (e.g., what queries are executing, what execution stage they are at) 
and the system state (e.g., resource utilization, data in cache).  
Crucially, the bi-directional LSTM architecture enables \modelname{} to separately understand the impact of queries submitted \textit{before} and \textit{after} a target query using the forward and backward passes.
%, which is crucial to \sysname{}.
%using the the forward and backward direction, respectively. 
The forward pass understands the impact of \textit{previous} queries, allowing \sysname{} to predict the \systime{} of an incoming query $Q_{n+1}$, given the queries $Q_1, ..., Q_n$ already running in the DBMS. 
In addition, for each running query $Q_i$, \sysname{} needs to understand its \systime{} change after submitting $Q_{n+1}$.
This can be estimated with the backward pass sending information from $Q_{n+1}$ back to each $Q_i$.
%Understanding the impact of \textit{subsequent} queries allows \sysname{} reason about the impact of submitting $Q_{n+1}$ on the runtimes of the already-running $Q_1, ..., Q_n$.
%\srm{I'm not sure I see what the latter case is looking forward -- it seems like you are predicting backwards again?}

%When ingesting queries online, \sysname{} cannot observe future queries and must predict a target query's runtime based only on queries submitted before it using \modelname{}'s forward pass.
%\sysname{} also needs to separately reason with the impact of this query on the existing running queries using \modelname{}'s backward pass.
%\ziniuw{Is this confusing? But I don't want to go through too many details on this.}

\begin{figure}[t]
    \centering
    \includegraphics[width=\columnwidth]{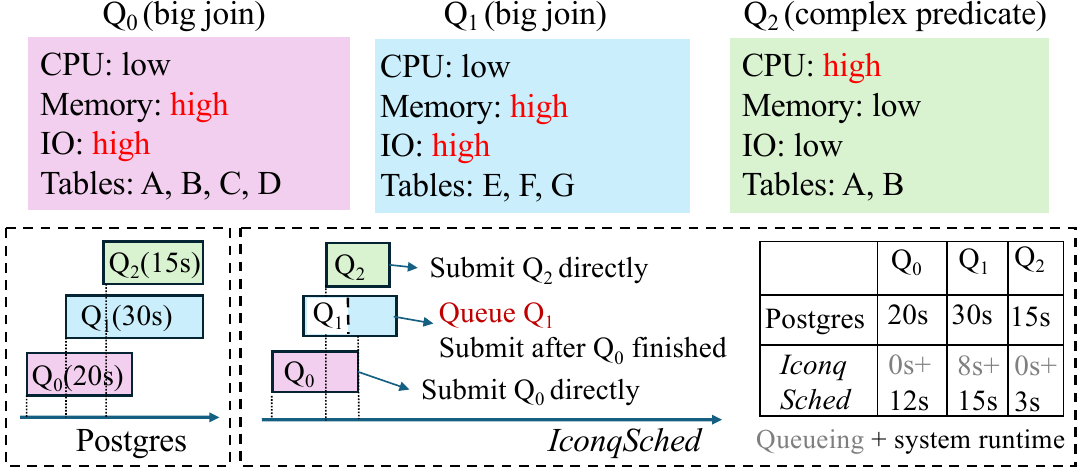}
    \vspace{-1em}
    \caption{Illustration of the operation of \sysname{}.}
     \vspace{-1em}
    \label{fig:demo}
\end{figure}

To address \textbf{Challenge 2}, we develop an efficient greedy scheduling algorithm to approximate the optimal solution.
At a high level, we first use \modelname{} to identify a set of candidate queries (among the queued queries) that are more beneficial to submit now than to leave in the queue.
Then, we use a scoring function based on \modelname{}'s predictions to calculate the overall \fulltime{} impact of submitting each candidate query.
The scheduling algorithm submits the best query as judged by the scoring function.
Immediately after, it re-evaluates the benefits of the remaining queued queries and iterates the above process to decide on the next best query, and so on.
The iterations stop when no queued queries are more beneficial to submit at the current decision timestamp, compared to staying in the queue.%\markos{a bit confusing - isn't it already the case that ``all waiting queries are considered'' even in the first pass it makes?}  

In Figure~\ref{fig:demo}, we illustrate  \sysname{} on a simple example with three queries. We summarize each query's resource requirements at the top of the figure, which are not directly observed but can be implicitly captured by \modelname{}. Left alone, the underlying DBMS (PostgreSQL~\cite{postgresql16}) will execute each query upon arrival, leading to high resource contention: executing $Q_0$ and $Q_1$ together results in high memory and I/O utilization and poor performance for both queries. 
\sysname{} instead determines that delaying the submission of $Q_1$ until $Q_0$ is finished would be beneficial, as it would minimize resource contention. It is worth noting that existing non-intrusive schedulers cannot accurately compute the \systime{} impact of deferring a query in order to make this decision. 
In the meantime, \sysname{} decides to execute $Q_2$ as soon as it arrives because the system state created by the execution of $Q_0$ is beneficial: resource usage is optimized because $Q_2$ and $Q_0$ need different resource types, while data sharing is maximized because the two queries access overlapping table sets.
As a result, \sysname{} achieves a $42\%$ reduction in the total \fulltime{}.

\sparagraph{Summary of Results.} We evaluate the performance of \sysname{} as a scheduler for two widely-used DBMSes, PostgreSQL~\cite{postgresql16} and AWS Redshift~\cite{redshift}. Our target metric is \fulltime{}---query queueing time plus \systime{}. With a few hours of training on query execution history, we can generate improved, non-intrusive scheduling decisions. On PostgreSQL, which has a simpler native scheduler and executor, \sysname{} has significant room to improve scheduling decisions: $16.2\% - 28.2\%$ in the mean and $33.6\% - 38.9\%$ in the tail (p-90) on realistic workload traces. On Redshift, which has a more sophisticated native scheduler (arguably state-of-the-art among commercial DBMS)~\cite{SaxenaRCLCCMKPN23, wu2024stage, nathan2024intelligent}, it is very hard for a non-intrusive scheduler to improve performance further. However, \sysname{} still improves performance by $10.3\% - 14.1\%$ in the mean and $14.9\% - 22.2\%$ in the tail.

\begin{figure*}[t]
	\centering
	\includegraphics[width=16.5cm]{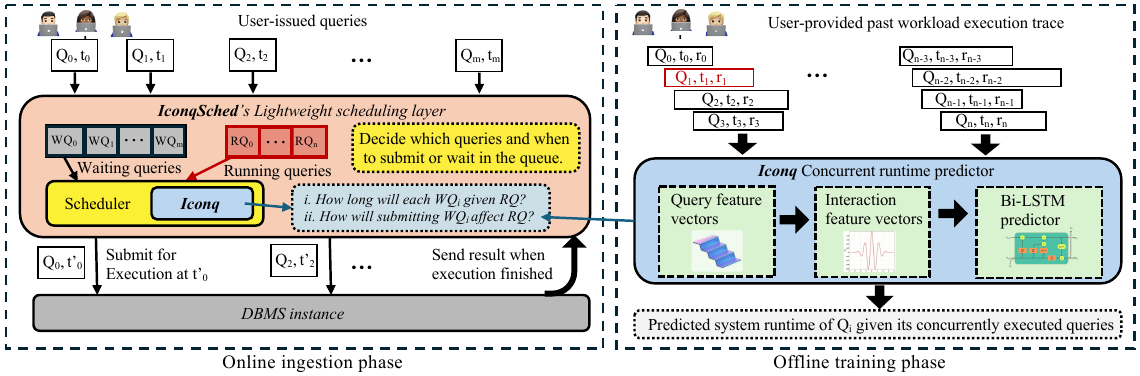}
    \vspace{-1em}
	\caption{System overview of \sysname{}.}
	\label{fig:overview}
\end{figure*}

In addition, we evaluated the performance of our \modelname{} predictor under various settings. \modelname{} achieved more accurate predictions than the baselines on real workload traces by $1.4\times - 2.4\times$ in the mean and $2.4\times - 4.9\times$ in the tail. We also show that \modelname{} is robust against changing workloads with varying numbers of concurrently executed queries and complex query templates. 
%It is worth noticing that \modelname{} could also benefit a wide range of other DBMS tasks that require accurate runtime estimations, such as query optimization~\cite{roy2000efficient} and maintaining SLA/SLO~\cite{ChiMHT11, MarcusP16}.

\smallskip
\noindent \textbf{Contributions.} In summary, we make the following contributions:

\begin{itemize}
    \item We propose \sysname{}, a novel non-intrusive query scheduler that can improve DBMS performance (Section~\ref{sec:system}). 
    \item We describe \modelname{}, an LSTM-based model that can accurately predict the \systime{} of concurrently executed queries (Section~\ref{sec:cost_model}).  \modelname{} can also independently benefit a wide range of other downstream DBMS tasks that require accurate \systime{} estimation, such as query optimization~\cite{roy2000efficient} and maintaining SLOs~\cite{ChiMHT11, MarcusP16}.
    \item We develop a greedy algorithm that uses \modelname{}'s predictions to make effective scheduling decisions (Section~\ref{sec:scheduling}).
    \item We conduct comprehensive experiments to evaluate the performance of both \sysname{} and \modelname{} (Section~\ref{sec:evaluation}). 
\end{itemize}

%% file: system.tex
\section{System Overview}
\label{sec:system}

%\sparagraph{Objective.}
In this work, we tackle the problem of non-intrusively ingesting and scheduling an online stream of OLAP queries $\{(Q_0, t_0), $ $\ldots$ $ (Q_m, t_m)\}$, where $t_i$ represents the arrival time of $Q_i$. We want to determine the optimal time to submit each query to the DBMS  in order to minimize total \fulltime{} as experienced by users, where \fulltime{} is query queuing time plus \systime{}, summed over all $Q_i$. 
%\srm{Does our work only apply to this average runtime case, or can it also be applied to fair scheduling or tail latency problems?} 
%\ziniuw{Our current implementation only applies to average runtime. I mention below that it is in principle easy to change for other objectives.}
Since most DBMSes do not preempt queries, this work focuses on non-preemptive scheduling. Our approach can also be naturally generalized to batch scheduling scenarios where all queries are ingested at the same time, or modified to minimize the median or tail query runtime. 

%\smallskip
\sparagraph{Architecture.} 
To address the non-intrusive scheduling problem effectively, \sysname{} is deployed as a \textit{lightweight scheduling layer} that sits outside the DBMS instance. This proxy layer design is commonly used for many systems~\cite{kraska2023check, yu2024blueprinting, RDSProxy, PgBouncer, ProxySQL} and can be generally applied to most DBMS~\cite{butrovich2023tigger}. Users can submit queries to this layer for better scheduling decisions, but we do not require every query to be submitted through \sysname{}. Instead, we only require information about all currently running queries to be available to \sysname{}. Many DBMS offer an interface to obtain a list of running queries, which can used to populate this information. % and the DB administrators who use our scheduling proxy can also access the submitted queries.

\sysname{} consists of two components, as shown in Figure \ref{fig:overview}: an \textbf{online scheduling algorithm} and a \textbf{concurrent query \systime{} predictor}, called \modelname{}. 
\modelname{} is trained offline on the user's executed workload trace. In the online ingestion phase, the scheduling algorithm invokes \modelname{} to make effective decisions. 
This offline-online two-phase design allows \sysname{} to offload a heavy amount of computation to the trained \systime{} predictor for efficient online decision-making. In the following, we will explain the two components of \sysname{} at a high level.
%and defer the details to Sections~\ref{sec:cost_model} and ~\ref{sec:scheduling}.

%Engineers can also implement this scheduling layer inside specific DBMSs to reduce overhead.

\smallskip
\sparagraph{Online scheduling algorithm:} 
As shown in Figure~\ref{fig:overview}, when ingesting queries online, \sysname{} first puts every arriving query in a queue and then uses \modelname{} to decide which queued queries to execute and when. Specifically, given $n$ queries running in the system $RQ$ = \{$RQ_1, \ldots, RQ_n$\} and $m$ queries waiting in the queue $WQ$ = \{$WQ_1, \ldots, WQ_m$\} at timestamp $t$, \sysname{} needs to decide which queries (if any) to submit to the system. Making the optimal decision is non-trivial because choosing queries to submit is an NP-complete problem~\cite{ullman1975np}, and deciding when to submit a query involves considering infinite possible submission timestamps. 
%The goal is to minimize the total end-to-end query runtime of $WQ$ and $RQ$, i.e., the sum of queuing time plus execution time for all $RQ_i$ and $WQ_j$.

Therefore, we design a greedy scheduling algorithm to approximate the optimal solution efficiently.
First, instead of considering infinite timestamps, our algorithm will only decide whether to submit a waiting query whenever a new query arrives or a running query finishes. 
Besides being efficient, this approach is also highly effective, because query arrival and completion highly impact the system and concurrent state, so we need to re-evaluate the benefits of submitting a waiting query or keeping it in the queue.
At each decision timestamp, \sysname{} uses a scoring function to evaluate these benefits accurately.
This scoring function uses \modelname{} to predict \textit{i) the \systime{} for each $WQ_i \in WQ$ given $RQ$} and \textit{ii) the total \systime{} change across $RQ$ if $WQ_i$ is submitted for execution}. 
Accordingly, \sysname{} identifies a set of candidate queries that are more beneficial to submit at the current decision timestamp, instead of later.
If this set is not empty, it selects the “best” query with the lowest score and submits it.
Then, it iterates this process to decide the next “best” query until no waiting queries are more beneficial to submit at the current decision timestamp.
This procedure avoids the exponential search space of candidate query sets.
We describe the details of this algorithm in Section~\ref{sec:scheduling}.

\sparagraph{Concurrent \systime{} predictor:} 
To make accurate \systime{} predictions for concurrent queries, \modelname{} first needs to observe a representative workload trace. 
%of executed queries (e.g., a few days of DBMS execution). 
In the offline training phase, we first analyze this trace to derive the concurrently executed queries for each query.
For example, for $Q_1$  (highlighted in Figure~\ref{fig:overview}), $\mathcal{Q}_1 =\{Q_0, Q_1, Q_2, Q_3\}$ are the concurrently executed queries  and $t_i$ and $r_i$ are its starting time and \systime{} (label).
Next, \modelname{} featurizes each query in $\mathcal{Q}_1$ into a \emph{\qvec{}} and uses the \qvec{}s to derive appropriate \emph{\tsvec{}s}.
%This featurization encodes a wide range of information on data access, query complexity, query start time, and estimated cost of operators and captures various complex query interactions between queries.
%Then, according to their submission time (i.e., $t_0, \ldots, t_3$), \modelname{} uses a time series encoding to represent the concurrent execution state of a target query.
Next, \modelname{} constructs a bi-directional LSTM~\cite{hochreiter1997long} to iteratively ingest these \tsvec{}s in the forward and backward directions. 
The LSTM model's \emph{hidden state} vector implicitly encodes the DBMS instance's system and concurrent state. 
After ingesting all four \tsvec{}s, the final hidden state will be used to predict the \systime{} of $Q_1$. We will discuss the details in Section~\ref{sec:cost_model}.

After training on the past workload execution history, \modelname{} can accurately estimate the \systime{} of an arbitrary query given any concurrent state.
Unlike prior work~\cite{wu2013towards, duggan2014contender, ahmad2011predicting, ahmad2009query, duggan2011performance, ahmad2011interaction, zhou2020query}, our design allows \modelname{} to capture complex query interactions and understand the changes in resource usage and data-sharing properties of the DBMS instance.
Therefore, \modelname{} can generalize to more complex query templates and a larger number of concurrent queries, which are not present in the training dataset (details in Section~\ref{subsec: exp-accuracy}).
Moreover, as explained in Section~\ref{sec:intro}, the bi-directional LSTM design allows \modelname{} to separately understand not only the impact of the already running queries $RQ$ on the \systime{} of $WQ_i$ but also the impact of submitting $WQ_i$ on the \systime{} of each $RQ_j$, which no existing approach can accurately capture.
%Therefore, \modelname{} can understand the runtime of a target query $WQ_i$ without knowing the future queries and the impact of $WQ_i$ on the running queries $RQ$, which is crucial to \sysname{}'s online scheduling process highlighted above.

\modelname{} treats the underlying DBMS engine as a black box, which generally applies to any DBMS engine with fixed on-premised hardware. For serverless and multi-tenant instances, a user's allocated resources may vary significantly depending on other users' resource usage, which is not observable by our model and cannot be accurately predicted. 
Exploring the \systime{} changes for different hardware types and serverless resources is an active field of research~\cite{mannino1988statistical, du1992query, bohm2000cost, negi2024pre} and is beyond the scope of this paper.

%% file: cost_mode.tex
\section{\modelname{} runtime predictor}
\label{sec:cost_model}

In this section, we explain the design of \modelname{}.  The goal of \modelname{} is to predict the \systime{} of a target query $Q$, given a set of concurrently executing queries $RQ = \{RQ_1, \ldots, RQ_n\}$.
At a high level, \modelname{} first featurizes $Q$ and each $RQ_i$  as \tsvec{}s and orders them according to their start timestamps.
Then, it uses an LSTM model to process the \tsvec{}s one at a time. It understands the concurrent state and system state changes as the LSTM model's hidden state updates. 
%\srm{This text, "one at a time" implies we have run the LSTM over each query each time a new query arrives?  Why can't we reuse the LSTM state from the last time we ran it? } \ziniuw{In the current implementation, we did not reuse the LSTM. So we run LSTM over each query each time a new query arrives. In principle, we can reuse the LSTM. However, the latency of LSTM is not a bottleneck in our experiment. I can explore other variants.} \srm{Maybe just say it could be optimized but is not a bottleneck in our current implementation?}
Finally, the LSTM model is used to predict $Q$'s \systime{}. 
We explain the details of our query featurization in Section~\ref{subsec: featurization} and predictive model design in Section~\ref{subsec: lstm}. 
In Section~\ref{subsec: training}, we explain the training and inference pipeline to make \modelname{} function inside \sysname{}.

\subsection{Query featurization}
\label{subsec: featurization}
We are given a target query $Q_j$ and a set of concurrently executing queries $\mathcal{Q}_j$, which includes all the queries the execution of which overlapped at least partially with $Q_j$. We featurize each query in $\mathcal{Q}_j$ into a \emph{\qvec{}} and then derive \emph{\tsvec{}s} to use as inputs to the LSTM model.

\begin{figure}[t]
	\centering
	\includegraphics[width=\columnwidth]{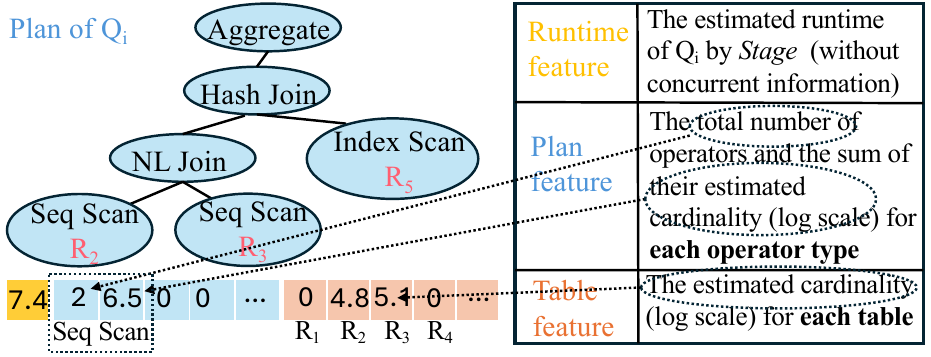}
 \vspace{-2em}
	\caption{\Qvec{}.
 %\srm{In this diagram, I would suggest putting labels under the blue and brown/orange boxes, with e.g., SS, HJ, AGG (for blue), and R1, R2, R3... (for brown/orange).  Also, for the plan features, might be good to show them as a tuple of count and runtime (e.g., \{\#:2,\showclock{0}{45}:6.5\} since it's not clear what the 6.5 represents.}
}
    \vspace{-1em}
	\label{fig:single_feature}
\end{figure}

\noindent \textbf{Representing a query:}
Many works have proposed featurization methods to represent a single query~\cite{kraska2023check, yu2024blueprinting, MarcusP19, SunL19, HilprechtB22}. 
In this work, we adapt the approach used by \textit{Stage}~\cite{wu2024stage} because of its proven effectiveness and efficiency inside a commercial DBMS. 
%\srm{say where stage is used?} An example resulting \qvec{} is shown in Figure~\ref{fig:single_feature}. \srm{Explain this figure}
Specifically, following \textit{Stage}'s method, we first identify $n_p$ physical operators that can significantly impact a query's \systime{} (e.g., scan, merge join, hash). We then initialize 2 \emph{plan features} in the \qvec{} for each of these operators: a count of how many times the operator appears in the plan of the query at hand and the total number of estimated rows (i.e., cardinality) processed by this operator, as shown in Figure~\ref{fig:single_feature}. 

One downside with the plan features derived from \textit{Stage}'s method is that they do not contain information about the tables accessed by a query, which is essential for understanding memory/buffer pool state and data share among concurrently executed queries. 
Therefore, we also add $n_t$ \emph{table features} to the \qvec{}, one for each of the  $n_t$ largest tables in the database. If a query touches one of these tables, we put an estimate of its cardinality for that table in the corresponding feature, as shown in Figure~\ref{fig:single_feature}.
%\srm{Are these the largest tables touched by the query or the largest tables in the database?  Clarify}\markos{tried to clarify}

Finally, we also add a \emph{runtime feature} with the average \systime{} of this query using the \textit{Stage} model, which is easy to train and can produce reliable estimates. Specifically, the \textit{Stage} model is able to estimate the  \systime{} of a single query without considering concurrent query information, as described in~\cite{wu2024stage}. 

As shown in Figure~\ref{fig:single_feature}, the resulting \qvec{} consists of the runtime feature, the $2*n_p$ plan features and the $n_t$ table features.
%\srm{instead of "single query feature" maybe call it a "query feature vector?}\markos{Yeah we can call the thing in Figure 3 the ``\qvec{}'' and the thing in Figure 4 the ``\tsvec{}''. I started implementing this change.} 
The hyperparameters $n_p$ and $n_t$ are tunable for each workload and database. In our evaluation, we set $n_p = 15$ and $n_t = 20$.

\begin{figure}[t]
	\centering
	\includegraphics[width=\columnwidth]{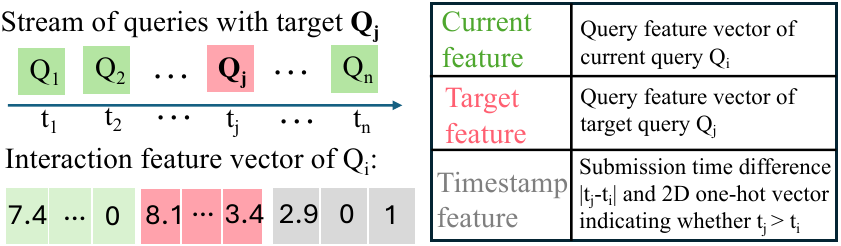}
	\caption{\TSvec{} with concurrent query execution information.}
  \vspace{-2em}
	\label{fig:concur_feature}
\end{figure}

\smallskip
\noindent \textbf{Representing a stream of queries:}
Assume a target query $Q_j$, for which we wish to estimate the \systime{}. Let the list  $\mathcal{Q}_j = \{Q_1, \ldots, Q_j, \ldots, Q_n\}$ be the   queries whose execution overlaps with that of $Q_j$, ordered by their submission times $t_1, \ldots, t_n$. 
%We want to encode this stream of queries as time series vectors that can be input to the LSTM model to predict the runtime of target query $Q_j$.
As shown in Figure~\ref{fig:concur_feature}, for each query $Q_i \in \mathcal{Q}_j$ (including the target query $Q_j$), we derive an \tsvec{} with three parts: i) the \qvec{} of $Q_i$, derived from the approach above, ii) the \qvec{} of $Q_j$, and iii) three \emph{timestamp features}. 
The three timestamp features are: i) the absolute difference $|t_i-t_j|$ between the submission times of $Q_i$ and $Q_j$, ii) an indicator for whether $t_i<t_j$, and iii) an indicator for whether $t_j<t_i$.

Concatenating the features of $Q_i$ and $Q_j$ helps the model understand their interactions, while the model can use the timestamp features to determine the extent of query execution overlap and infer the execution stage each query is in. 
We denote the \tsvec{} for $Q_i$ as $x_i$ for each query in $Q_i \in \mathcal{Q}_j$ and derive a time series of \tsvec{}s $\{x_1, \ldots, x_n\}$.

\subsection{Modeling concurrent query execution}
\label{subsec: lstm}

Recall that the goal of \modelname{} is to accurately predict the \systime{} of target query $Q_j$ when concurrently executing with  queries $\mathcal{Q}_j = \{Q_1, \ldots, Q_n\}$, by understanding the impact of concurrency and system state changes (e.g., resource utilization).
\modelname{} first uses the above approach to derive a time series of \tsvec{}s $\{x_1, \ldots, x_n\}$. Then, \modelname{} ingests these vectors into a bi-directional LSTM model, which updates its internal hidden state to models the DBMS instance's system state changes. 

Long Short-Term Memory (LSTM)~\cite{hochreiter1997long} is a type of recurrent neural network (RNN) architecture that is specifically designed to process and remember sequences of data (e.g., time series, speech) and achieves the state-of-the-art performance in various fields~\cite{sutskever2014sequence, jozefowicz2015empirical, lipton2015critical}.
An LSTM model processes a time series of inputs one at a time and uses its internal hidden state to store an evolving summary of the processed information from the previous inputs.
At each step, the model decides what information to discard, understands the dependencies between the input and the previous hidden state, and updates the hidden state according to the latest information. 
We design \modelname{} based on a special type of LSTM, a bi-directional LSTM model~\cite{schuster1997bidirectional}.
It processes the time series of inputs in two directions, forward and backward, and uses two hidden states to understand the dependencies from past and future inputs. 

\begin{figure}[t]
	\centering
	\includegraphics[width=\columnwidth]{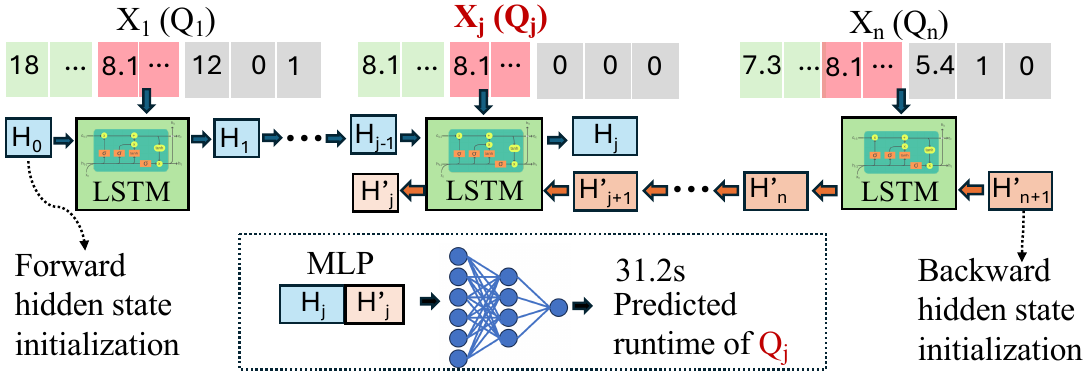}
     \vspace{-2em}
	\caption{Concurrent query runtime predictor. }
	\label{fig:lstm}
\end{figure}

%\srm{This discussion is hard to follow because it's not clear what the different LSTM gates do -- is this a unique utilization of LSTMs?  Can we somehow say how LSTMs are being used in an unusual way here? For statements like "the forward hidden state will implicitly capture the concurrency and system state information after submitting Q\_i" it's not really clear to me *how* it captures this } \ziniuw{We did not use unique utilization of LSTMs. I will explain how LSTM works in high-level with non-deep-learning jargon and focus more on why is its mechanism aligned with the problem that we wanted to solve.}
%\srm{I tried to clarify that this is the inference process, might be good to say somewhere that we describe how we use the model first and then how we train it below?}
%\ziniuw{Yes, the training process is discussed in section 3.3}
We describe how \modelname{} predicts the system runtime of a target query $Q_i$ given interaction fracture vectors of its concurrently executed queries in Figure~\ref{fig:lstm}.
%The (inference) process \modelname{} used for runtime prediction is shown in .
\modelname{} takes as input the time series $\{x_1, \ldots, x_n\}$  and the index $j$ of the target query. It then traverses the time series in both the forward and backward directions.
In the forward pass, \modelname{} initializes the forward hidden state $H_0$ as a zero matrix and ingests the time series in the forward direction.
%, and then the \tsvec{}s $\{x_1,..., x_j\}$ are ingested one at a time.
Upon ingesting each $x_i$, \modelname{} updates its previous forward hidden state $H_{i-1}$ and produces $H_i$.
After ingesting $x_j$, the forward hidden state $H_j$ is saved, which captures the impact of the preceding queries $Q_1, \ldots, Q_{j-1}$ on $Q_j$.
In the backward pass, the backward hidden state $H'_{n+1}$ is similarly initialized, and then the \tsvec{}s $\{x_j,..., x_n\}$ are ingested one at a time \emph{in reverse order}, with the ingestion of $x_i$ producing backward hidden state $H'_i$. After ingesting $x_j$, the backward hidden state $H'_j$ is saved, which captures the impact of the subsequent queries $Q_{j+1}, \ldots, Q_n$ on $Q_j$.
When both the forward and the backward passes are complete, \modelname{} updates its final hidden state as a concatenation of the forward hidden state $H_j$ and the backward hidden state $H'_j$.
It then passes the final hidden state through a multi-layer perception (MLP) to predict the \systime{} of $Q_j$.
%We then pass the final hidden state through a multi-layer perception (MLP) to predict the \systime{} of $Q_j$.

%For each input $x_i$, \modelname{} uses the three LSTM gating mechanisms—input, forget, and output gates—to process the input, discard outdated information, and update the forward hidden state to $H_i$, maintaining long-term dependencies while filtering out irrelevant details. The forward hidden state $H_i$ will implicitly capture the concurrency and system state information after submitting $Q_i$, such as CPU/memory utilization, data in the buffer pool/cache, and resource need for the existing concurrent queries.  After ingesting $x_j$, the forward hidden state $H_j$ is saved, which captures the impact of the preceding queries $Q_1, \ldots, Q_{j-1}$ on $Q_j$.

%When both forward and backward passes are complete, \modelname{} will have accounted for the impact of all concurrent queries in $\mathcal{Q}_j$ on $Q_j$ and updated its final hidden state as a concatenation of the forward hidden state $H_j$ and the backward hidden state $H'_j$.
%We then pass the final hidden state through a multi-layer perception (MLP) to predict the \systime{} of $Q_j$.

\smallskip
\noindent \textbf{Analysis and discussion:} The design of \modelname{} as a bidirectional LSTM\cite{hochreiter1997long} is well-suited for the task of \systime{} prediction in three core ways. Whereas, the state-of-the-art deep learning architectures, e.g. transformer~\cite{vaswani2017attention}, do not have such properties.

First, \textbf{LSTMs explicitly update a hidden state}. This is important because concurrent queries do not impact each other's \systime{}s directly; they do so by impacting the state of the DBMS (things like the memory/CPU usage or buffer pool state), which in turn can have a positive or negative effect on \systime{}. Our choice of an LSTM model, where each $x_i$ is submitted one at a time and updates the hidden state to $H_i$ before processing $x_{i+1}$, naturally parallels this reality. This hidden state implicitly captures the state of the DBMS that it is able to learn from the training data; for example, if the training data includes concurrent queries with large joins over the same table that run very slowly, the hidden state can learn to reflect this type of interaction and model its effect on other concurrent queries. Because our featurization captures the high-level properties of the queries (e.g., cardinalities and plan structure) rather than using database- or table-specific features (e.g., column names), this learning is generalizable to unseen feature patterns, such as unseen query templates (see Section~\ref{subsec: exp-accuracy}).

%For example, if several queries with large estimated cardinality for the join operators (captured in the \qvec{} of Figure~\ref{fig:single_feature})  are submitted in a short time interval, we expect them to have long runtimes because of high memory contention.  \modelname{} can recognize the pattern in the features of these queries, understand their interactions, and associate the large estimated cardinalities with high memory/CPU utilization and their long system runtime.

%As another example \markos{do we need a second example?} \ziniuw{probably not I guess} \markos{we could remove it if we need space}, consider a query with a large cardinality for the join operators (high memory utilization) concurrently running with a query with a large cardinality for the scan operators (high IO utilization). These two queries do not significantly impact each other because they prioritize different types of resources.  After training, \modelname{} can associate patterns in the features with complex query interactions and resource utilization.

Second, \textbf{LSTMs selectively forget information over time}. This is desirable because parts of the DBMS state are affected by each query for different durations - evaluating a complex predicate will create a temporary demand for CPU cycles, but bringing a table into the buffer pool will have a longer-lasting impact. During its training, the LSTM learns \emph{which} features to ``remember'' in its hidden state and \emph{for how long}. Concretely, \modelname{} can learn that the features corresponding to some finished operators have little effect on other concurrent queries and discard them from its hidden states. 
%\ziniuw{I will come up with a more concrete example here, otherwise it sounds a bit hand-wavy.}
%Moreover, the LSTM model learns to filter out unimportant information while preserving relevant information in its hidden state~\cite{hochreiter1997long}.
%For example, \modelname{} can learn that the feature corresponding to some operator has little effect on other concurrent queries after some time. 
This selective forgetfulness means that the hidden state remains informative even as  
the time series $\{x_1, ..., x_n\}$ grows longer, allowing \modelname{} to generalize to scenarios with a large number of concurrent queries (see Section~\ref{subsec: exp-accuracy}). This observation is similar to the situation where an LSTM trained on short sentences can generalize to a longer corpus of text~\cite{hochreiter1997long, karpathy2015visualizing}.

%This ability of LSTMs to capture hidden states and discard unimportant information allows \modelname{} to generalize to unseen scenarios with more concurrently executing queries 

%the LSTM hidden state functions like a working memory that stores relevant information. Thus, \modelname{} can use the forward and backward hidden states to capture information about the concurrently executing queries $\mathcal{Q}_j$ that may affect the runtime of $\mathcal{Q}_j$. Similar to the examples above, these hidden states implicitly capture the system state changes (e.g., memory, I/O, bandwidth, CPU utilization, and data in buffer pool/cache). This information directly affects \systime{}. 

Third, \textbf{a bidirectional LSTM understands the impact of the past and the future separately}. This is important because
not only is the \systime{} of a newly-submitted query $Q$ affected by the current state of the DMBS, created by already-running queries $RQ$ (the ``past''); $Q$ also itself affects the state of the DBMS once submitted, possibly impacting the \systime{} of each query in $RQ$ (from the perspective of which, such impact comes from the ``future''). For \sysname{} to make good scheduling decisions,
each of these effects must be understood and considered (see Section~\ref{subsec: training}).

%\srm{This is a little confusing because this memory IO bandwidth etc are not actually given to the model -- we are assuming that it is able to model this.  Maybe say "The hidden state implicitly captures the state of the DBMS that it is able to learn from the training data;  for example, if the training data includes concurrent queries with large joins over the same table that run very slowly, the hidden state can reflect this type of interaction and model its effect on other concurrent queries." }
%\ziniuw{yes, I agree this claim can not be backed up. But your example sounds very similar to the examples above. The thing is LSTM is like a black-box and its hidden state is definitely doing something clever. So we tried to give this black box a reasonable explanation (seems weird). Should we completely remove this part?} \markos{rewrote the ``analysis and discussion'' part and moved these comments to the end. lmk what you think} \ziniuw{I like it and I edited it a bit.}

\subsection{Offline training and Online Inference}
\label{subsec: training}
In the following, we describe how we train \modelname{} in the offline phase and how we use it to make inferences in the online phase.  

\smallskip
\noindent \textbf{Offline training:} 
In this work, we assume that users  provide us with a query execution history in the form of a workload trace (e.g., a few days of DBMS execution). 
%\chunwei{The cardinality of the query set matters a lot. If the number of distinct queries is too small, the prediction can be easy but less meaningful.} \ziniuw{Yes. But we observed a similar workload so if the cardinality is small, we expect the users to submit similar queries. We can not expect our model to work for scenarios like users only execute two types of queries in the past but suddenly have thousands queries after deploying our model.}
This is reasonable because we hypothesize that the users will deploy \sysname{} on an existing workload on a particular DBMS instance. 
In the case of a new DBMS instance, \sysname{} needs to collect data for a few days before making effective scheduling decisions. 
We plan to address this ``cold-start'' problem in future work.

The user-provided training trace needs to contain a list of queries along with their query plans, submission time, and \systime{}. 
First, we train a \textit{Stage} model from these queries to predict the average \systime{} per query, without considering the impact of concurrent execution~\cite{wu2024stage}. 
These predictions form the \emph{runtime feature} of the \qvec{}, as mentioned in Section~\ref{subsec: featurization}.
Then, for each query $Q_j$, we identify the list $\mathcal{Q}_j$ of overlapping queries and compute the \tsvec{}s, as explained in Section~\ref{subsec: featurization}.
Each $Q_j$ and its associated time series of \tsvec{}s are forwarded to the bi-directional LSTM model to predict its \systime{}, per Section~\ref{subsec: lstm}. 
The LSTM model is trained end-to-end using L1 loss and Q-loss~\cite{kipf2018learned}.
%When users' data and workload drastically change, our model needs to be retrained based on the newly observed executions.
We do assume that the underlying hardware the DBMS runs on does not change frequently --  \modelname{} needs to be re-trained if the user switches to a new hardware type. 
Thus, the current version of \modelname{} is not eligible for serverless settings where virtualization hardware may have a high-performance variability.
%We left exploration in this direction as future work.
%\chunwei{Seems to be not eligible for serverless setting. And cloud services also have the resource variances. } 

\smallskip
\noindent \textbf{Online inference:} 
\modelname{} can be used for a wide range of downstream DBMS tasks that require accurate \systime{} estimation, such as optimization~\cite{roy2000efficient} and maintaining SLA/SLOs~\cite{ChiMHT11, MarcusP16}. In \sysname{}, it is used to support our online scheduler in deciding which queries from the waiting queue $WQ = \{WQ_1, \ldots, WQ_m\}$ to submit for execution and when to submit them (see Section~\ref{sec:scheduling}).
Specifically, \sysname{} needs to know i)  the runtime of $WQ_i$ if it is submitted at timestamp $t$, given the currently running queries $RQ  = \{RQ_1, \ldots, RQ_n\}$; and ii) how the runtime of each running query $RQ_j \in RQ$ will change if $WQ_i$ is submitted. 

To answer the first question, we consider $WQ_i$ as the target query and the running queries $RQ$ as its concurrently executed queries. We derive a time series of \tsvec{}s as per Section~\ref{subsec: featurization} and input them into \modelname{}. \sysname{} \textit{only needs to perform the forward pass} on this time series since no query has been submitted after the target $WQ_i$. Note that the estimated \systime{} we obtain for $WQ_i$ this way does not account for any future queries that may possibly be submitted while $WQ_i$ is running. If \sysname{} does decide to submit such queries, their impact on $WQ_i$ will be considered at that time, at which $WQ_i$ will be one of the ``running queries''.
%it can adjust the predicted runtime with LSTM backward pass and answer the second question.
This design enables \sysname{} to make robust decisions given imperfect information, i.e., without knowing what queries will arrive. For implementation efficiency, we only need to featurize each query once when ingesting it and cache its featurization for later use.

To answer the second question, for each query $RQ_j \in RQ$, we re-predict its \systime{} to account for the impact of query $WQ_i$. Specifically, for each $RQ_j$, let \{$x_0, \ldots x_n$\} be its time series of \tsvec{}s. \modelname{} appends another element to this series: the \tsvec{} $x_{n+1}$ for the interaction of $WQ_i$ with the target $RQ_j$.
%feature (as illustrated in Figure~\ref{fig:concur_feature}). \modelname{} will featurize the single query feature of $WQ_i$ (as illustrated in Figure~\ref{fig:single_feature} with $RQ_j$ being the target query) and add this feature as $x_{n+1}$ to $RQ_j$'s time series feature.
Then, \modelname{} will
input \{$x_0, \ldots x_{n+1}$\} into the LSTM model to predict the \systime{} of $RQ_j$. $WQ_i$ only affects the backward pass of each $RQ_j$'s \systime{} prediction because it is submitted after $RQ_j$.

%Predicting queries arriving in the future is an orthogonal field of research~\cite{ma2018query,huang2024sibyl}. We will explore integrating a workload forecaster with our scheduler in future work.

%% file: scheduling.tex
\section{Scheduling algorithm}
\label{sec:scheduling}

In this section we describe how we use \modelname{} to make scheduling decisions when ingesting an online stream of OLAP queries. 
Specifically, given $n$ queries running on the system $RQ$ = \{$RQ_1, \ldots, RQ_n$\} and $m$ queries waiting in the queue $WQ$ = \{$WQ_1, \ldots, WQ_m$\}, \sysname{} needs to decide which queries (if any) in $WQ$ to submit for execution to minimize the total query \fulltime{} (\textit{e2e-time}), i.e., the sum of queuing time plus \systime{} for all queries in $RQ$ and $WQ$.

Briefly, \sysname{} invokes the scheduler whenever a query arrives, is submitted, or finishes. At each invocation, \sysname{} uses \modelname{} to select a set of \emph{candidate queries} from $WQ$, each of which the model determines would benefit from being submitted now.  
\sysname{} then scores each candidate query using a scoring function and submits the one with the highest score.  In the following, we will provide details on when to invoke the scheduler, how to select candidate queries, how to score them, and how to optimize the submission of short-running queries. We provide the pseudocode of the overall \sysname{} scheduling algorithm in Appendix~\ref{sec:app-algo}.

\smallskip
\noindent \textbf{Invoking the scheduler:} 
At each invocation, \sysname{} decides the best waiting query to submit right now, if any.
%If \sysname{} scheduled queries to be submitted in the future, it would need to make assumptions about the (absence of) query arrivals in the meantime, possibly hurting robustness. \ziniuw{This sentence is a bit unclear. The two things are equivalent. We can just say the next sentence right?} \markos{sure yeah we can remove the previous sentence} Instead, \textbf{}. 
This leaves the problem of when exactly each invocation of \sysname{} should occur.  
A naive approach would invoke the scheduler at equally spaced times (e.g., every $t_{sched} = 5$ seconds). 
However, this approach has drawbacks. First, it may lead to long queuing times for queries with a \systime{} much smaller than $t_{sched}$, which happen to arrive at inopportune moments, hurting their \fulltime{}. 
Second, this approach would be computationally wasteful during periods of no significant system state changes (e.g., if the system is idle or a very long query is running and dominating the resources). To avoid these drawbacks, we instead invoke \sysname{} based on \emph{events} rather than periodically. In particular, we invoke \sysname{} i) each time a new query \emph{arrives} in the queue and ii) each time a running query \emph{finishes} its execution.

\smallskip
\noindent \textbf{Selecting candidate queries:}
At each invocation, for each queued query $WQ_i \in WQ$, \sysname{} computes the benefit of submitting $WQ_i$ for execution \emph{now}, against submitting it \emph{at one of $L$ future timestamps}, where $L$ (for ``lookahead'') is a hyperparameter. Note that \sysname{} can only submit a query whenever it is invoked; therefore, ``at one of $L$ future timestamps'' must mean \emph{at one of the $L$  future invocations of \sysname{}}. 
As described above, future invocations of \sysname{} will happen in response to a query arriving or finishing its execution. 
Recall that \sysname{} does not assume any knowledge about future \emph{query arrivals}, the prediction of which is an orthogonal field of research~\cite{ma2018query,huang2024sibyl}. 
Therefore, we only compute the benefit of submitting $WQ_i$ for execution \emph{now}, against submitting it at one of $L$  future invocations of \sysname{} \emph{because of completed query execution}.

%We first define a tunable hyperparameter, $lookahead$, representing the number of future timestamps to compare against for the queued queries.  \srm{What is lookahead?  A number of seconds?  How do you choose which future timestamps to look at if you don't know when future queries will arrive???}  \ziniuw{we defined it here: "the number of future timestamps to compare against". We assume no query will arrive, and predict the finishing time of existing queries} \srm{Sorry I'm being dense but I don't get it -- if we don't model when future queries will arrive how do we know what the future timestamps are that we should compare against?} \markos{Let me take an hour to rewrite this until 2pm to make it clearer}

%; we plan to explore integrating a forecaster with \sysname{} as future work.
%Thus, if \sysname{} does not submit $WQ_i$ now, it will reconsider $WQ_i$'s submission only when one of the running queries finishes (i.e., at the next invocation). 

We use \modelname{} to predict when each running query in $RQ$ will finish, sort the finish times in increasing order, and take the $L$ soonest ones, $t_1, \ldots, t_{L}$, where $t_l$ corresponds to the $l$-th soonest time at which a running query is expected to finish. Thus, \sysname{} needs to compare the benefit of submitting $WQ_i$ now (at $t$) or in the future at any  $t_l$. This benefit has two components: i) $\delta_1(t_l)$, measuring how much longer the \fulltime{} of $WQ_i$ will be if submitted at $t$ instead of at $t_l$; and ii) $\delta_2(t_l)$, measuring how much longer the \fulltime{} of each query in $RQ$ will be  if $WQ_i$ is submitted at $t$ instead of at $t_l$.

\vspace{-1em}
\begin{align}
    \label{equ: sched-delta1}
    \delta_1(t_l) = \mathbf{Iconq}&(WQ_i, RQ, t) - (  \mathbf{Iconq}(WQ_i, RQ', t_l) + t_l - t) \\
    \label{equ: sched-delta2}
    \delta_2(t_l) = \sum_{j} \ \ & \mathbf{Iconq}(RQ_j,  \{WQ_i\} \cup RQ, t) \ - \ \nonumber \\
    &\mathbf{Iconq}(RQ_j,  \{WQ_i\} \cup RQ', t_l) 
\end{align}
\vspace{-1em}

As shown in Equation~\ref{equ: sched-delta1}, $\delta_1(t_l)$ is the difference in the predicted \fulltime{} of $WQ_i$ if submitted at $t$ concurrently with $RQ$ (i.e. \systime{} $\mathbf{Iconq}(WQ_i, RQ, t)$ plus no queuing time), or at $t_l$ concurrently with $RQ'$ (i.e., \systime{} $\mathbf{Iconq}(WQ_i, RQ, t_l)$, plus extra queuing time $t_l - t$), where $RQ'$ excludes the queries in $RQ$ that will have finished by $t_l$.
%\markos{shouldn't we have removed from $RQ$ the queries that finished?} \ziniuw{yes, we have that in appendix, but too complicated to explain here and is a minor point.}
Similarly, as shown in Equation~\ref{equ: sched-delta2}, $\delta_2(t_l)$ is the sum over all $RQ$ of the differences in their \fulltime{}s if $WQ_i$ is submitted at $t$ or at $t_l$. 
Note that the $RQ$ have already incurred their queuing time, so the change in their \systime{} will be the same as in their \fulltime{}.
%\ziniuw{the last sentence is obvious I assume. remove?} \markos{It is a bit obvious, but I added it to avoid confusion. I will shorten it.}

%\textit{e2e-time} of $WQ_i$ between submitting it at $t$ and at $t_k$, denoted as $\delta_1$ in . Here, $\mathbf{Iconq}(WQ_i, RQ, t)$ refers to \modelname{}'s predicted runtime of $WQ_i$ submitted at timestamp $t$ given $RQ$ running queries and $t_k - t$ refers to the additional queueing time of $WQ_i$.
%Then, we calculate the impact of submitting $WQ_i$ at timestamp $t$ and $t_k$ on each running query $RQ_j$. We sum the impact difference on all running queries $RQ$ as $\delta_2$ in Equation~\ref{equ: sched-delta2}.
Therefore, $\delta_1(t_l) + \delta_2(t_l) > 0$ for some $t_l$ suggests that submitting $WQ_i$ at $t_l$ is more beneficial than submitting now, so %at $t$, i.e., the sum of total \textit{e2e-time} for all $RQ$ and $WQ$ is smaller for submitting $WQ_i$ at $t_k$. 
we should keep $WQ_i$ in the waiting queue. 
Otherwise, $\delta_1(t_l) + \delta_2(t_l) \leq 0$ for all $t_l$ suggests that it is more beneficial to submit $WQ_i$ now. In this case, we will consider $WQ_i$ as a candidate query for submission. \sysname{} will iterate the above process for all waiting queries in $WQ$ and derive a selected set of candidate queries.

Our selection criterion is inspired by works using cost-based scheduling (CBS) for maintaining SLOs~\cite{chi2013distribution, peha1991scheduling}, where they define a criterion to compute the expected benefits of delaying a query. However, their approach cannot account for query interactions in concurrent execution.
The hyperparameter $L$ balances the algorithm's efficiency and accuracy. A larger $L$ value will consider more possible submission times and better justify the decision to submit a query now. However, it will impose a larger computation overhead and it may be less reliable, since it is more likely that new queries will arrive and \sysname{} is invoked in the meantime. In our evaluation, we set $L = 2$ and verified that a larger value does not improve the scheduler's performance. 

\smallskip
\noindent \textbf{Scoring candidate queries:}
Submitting all candidate queries together is not optimal because we have not accounted for their interactions \emph{with each other}. 
For example, submitting any of $WQ_1$ or $WQ_2$ now may be beneficial, but after submitting $WQ_1$, \sysname{} may find that deferring $WQ_2$ is better because of their negative interference.
Since evaluating an exponential number of query interactions in the candidate query set is expensive, \sysname{} uses a greedy algorithm to approximate the optimal solution. 
At a high level, \sysname{} scores the candidate queries, ranks them and selects \emph{one query} to submit for execution, to avoid the huge decision space. Submitting this query will immediately trigger a new invocation of \sysname{}, at which point the candidate queries will be re-derived and the scoring function re-evaluated. 

We design the scoring function of \sysname{} to minimize the total \fulltime{} for all $RQ$ and $WQ$.
For each candidate query $WQ_i$, the score is a sum of three components: i) $\Delta_1(WQ_i)$, measuring how promising the current system state is for $WQ_i$'s execution, ii) $\Delta_2(WQ_i)$, measuring how badly the running queries $RQ$ will be affected by $WQ_i$, and iii) how long $WQ_i$ has been in the queue for. Smaller scores are better. 

\vspace{-1em}
\begin{align}
    \label{equ: score-delta1}
    \Delta_1(WQ_i) &= \mathbf{Iconq}(WQ_i, RQ, t) - \mathbf{Stage}(WQ_i) \\
    \label{equ: score-delta2}
    \Delta_2(WQ_i) &= \sum_{j} \  \mathbf{Iconq}(RQ_j,  \{WQ_i\} \cup RQ, t)  \ - \ \nonumber \\
    &\quad \quad \quad \mathbf{Iconq}(RQ_j, RQ, t) \\
    \label{equ: score}
    score &= \Delta_1 + \Delta_2 - \lambda * Queueing\_time(WQ_i)
\end{align}
\vspace{-1em}

As shown in Equation~\ref{equ: score-delta1}, $\Delta_1(WQ_i)$ is the difference between i) $WQ_i$'s predicted \systime{} if submitted at time $t$ with queries $RQ$ also running and ii) $WQ_i$'s average runtime, as predicted by \textit{Stage} (see Section~\ref{subsec: featurization}). Estimating this is necessary because \sysname{} does not reject or preempt queries, so $WQ_i$ will eventually be executed at some point, so we want to identify especially opportune moments for executing it. 
We use $WQ_i$'s average runtime to approximate its runtime when submitted at an arbitrary future point. A negative value of $\Delta_1(WQ_i)$ suggests that the current system state has a positive impact on $WQ_i$ that makes its \systime{} shorter than average. Since both terms of this equation consider the immediate submission of $WQ_i$, the predicted \systime{} impact is also the predicted \fulltime{} impact.

As shown in Equation~\ref{equ: score-delta2}, $\Delta_2(WQ_i)$ is the sum over all $RQ$ of the differences in their \fulltime{}s if $WQ_i$ is submitted now (at $t$) or not at all. Again, the $RQ$ are already submitted, so computing the \systime{} impact is equivalent to computing the \fulltime{} impact. Thus, $\Delta_1(WQ_i) + \Delta_2(WQ_i)$ corresponds to the total query \fulltime{} impact if we submit $WQ_i$ now.
Finally, as shown in Equation~\ref{equ: score}, the overall scoring function also incorporates a penalty for queries that have been queued for too long, controlled by a a hyperparameter $\lambda$, in order to prevent a query from starving in the waiting queue. It is possible to modify the scoring function to optimize the median or tail runtime of all queries~\cite{koenker2005quantile, ben2009robust}. We leave this exploration as future work.

The candidate query with the smallest score is submitted for execution, and then we immediately trigger another invocation of \sysname{}, at which point the candidate queries will be re-derived, the scoring function re-evaluated, and the next best query submitted. This greedy algorithm will iteratively select one query to submit per invocation until some invocation derives an empty candidate set (at most $|WQ|$ invocations). 

\smallskip
\noindent \textbf{Short query optimization:}
\sysname{} uses its trained \textit{Stage} predictor to predict the average runtime of a query upon its arrival in the queue. \sysname{} will directly submit this query if its average runtime is less than $\tau$ seconds, before even deriving the candidate set. 
We do this because scheduling has a limited impact on short-running queries -- the runtime of short-running queries is very stable under different system loads and concurrent states because they use very limited resources (prior work made similar observations~\cite{ahmad2008modeling, ahmad2011interaction}).
Another reason is that \sysname{} needs to invoke \modelname{} to make scheduling decisions, incurring an additional $10-100ms$ of overhead, which can be significant for short-running queries. In our evaluation, we set $\tau$ to $5$ seconds.
%Therefore, bypassing \sysname{}'s decision logic and directly submitting short-running queries is a reasonable design.

% \textit{Stage} has been shown to be accurate and efficient in Amazon Redshift~\cite{wu2024stage}. \markos{we should put this sentence at the point where we mention Stage for the very first time}

\smallskip
\noindent \textbf{Analysis:}
Each invocation of \sysname{} calls \modelname{} O($L * |RQ| * |WQ|$) times in a batch.
The batch inference significantly accelerates the inference process. 
Specifically, for each $t_l$, $l\in[1, L]$, we need to predict the impact of each $WQ_i \in WQ$ on each of the running queries $RQ$. For most of the OLAP workloads~\cite{van2024tpc, snowflake-nsdi20}, the number of concurrently running queries $|RQ|$ is relatively small (e.g., $<10$).
Compared to the heuristic-based schedulers~\cite{SaxenaRCLCCMKPN23, gupta2009fair, mehta2008bi, ahmad2011interaction}, \sysname{} uses a scoring function that can accurately evaluate the benefits of queries and presents a %principled \srm{What do we mean by principled?} 
framework that can make more effective scheduling decisions. 
%In addition, \sysname{} can support incrementally adding user-defined rules. \markos{like what} %\srm{What do we mean by "explainable"?  LSTMS are pretty black box} 
While decision-making may be suboptimal since our greedy algorithm cannot account for all query interactions in $WQ$, we find experimentally that \sysname{} results in significant reduction in \fulltime{} versus existing commercial schedulers.

%% file: experiment.tex
\section{Evaluation}
\label{sec:evaluation}
In this section, we first describe our experimental setup in Section~\ref{subsec: exp-setup}. Then, we address the following questions:

\smallskip
\noindent $\bullet$
How much does our scheduler improve the query performance of existing DBMSs on a practical workload (Section~\ref{subsec: exp-e2e})?

\smallskip
\noindent $\bullet$
How does our scheduler behave with a varied number of clients hitting the system (Section~\ref{subsec: exp-adaptivity})?

\smallskip
\noindent $\bullet$
What are the accuracy and overhead of our runtime estimator? How robust is it under a changing workload (Section~\ref{subsec: exp-accuracy})?

\subsection{Experimental Setup}
\label{subsec: exp-setup}

In this section, we explain our experimental environment, the baselines we compare with, and the workloads.

\sparagraph{Environment:} We note that \sysname{} is not tailored to a particular DBMS but can be applied to a wide range of systems. We conduct experiments on two DBMS: Postgres~\cite{postgresql16} and Redshift~\cite{redshift}. 

We chose Postgres because it is one of the most widely used open-source DBMSs with countless applications built upon it~\cite{dbengines2024}. However, Postgres has a very simple scheduling approach that uses the multi-programming level (MPL) to control the maximum number of concurrent queries in an instance and executes queries in a first-in-first-out fashion. Thus, Postgres leaves us room to improve its scheduling decisions in a non-intrusive way. Our experiments use AWS ``db.m5.xlarge'' Postgres 16.2 instances with 4 vCPUs, 16GB RAM, 1000GB storage, and 2000 provisioned IOPS.

We chose Redshift because it is one of the most widely used commercial data warehouses dedicated to OLAP workloads~\cite{dbengines2024, SaxenaRCLCCMKPN23}, which are the focus of this work. 
Redshift contains a state-of-the-art workload manager that can make effective decisions on admission control, scheduling, and resource allocation~\cite{wu2024stage, SaxenaRCLCCMKPN23, nathan2024intelligent}. Thus, it is very challenging for non-intrusive schedulers to improve Redshift. Our experiment uses AWS ``dc2.large'' Redshift instances with 1 node, 2 vCPUs, 7.5 GB of RAM, and 160 GB of SSD.
%\markos{add citation}

We tuned the configurations (e.g., MPL) of both the Postgres and Redshift instances on the workload traces. Our scheduler and all baselines are trained and served on a Linux machine with 40 Intel(R) Xeon(R) Gold 6230 CPUs and 128GB RAM.
%\markos{maybe explian a bit more how the client/proxy is set up?} \ziniuw{maybe, but there is nothing smart at the point. Now I simulated users rather than having an actual connection like Brad. }

%\smallskip
\sparagraph{Scheduling Baselines:} We compared \sysname{} to the following non-intrusive query scheduler baselines:

$\bullet$ \textit{Postgres/Redshift}: the original DBMS with tuned knobs.

$\bullet$ \textit{PGM}: the PGM scheduler~\cite{mehta2008bi} that estimates the memory usage for each query. The PGM scheduler uses the estimated memory as admission control to keep the total memory usage of concurrently executed queries under the total memory of the instance. Whenever admitting a query would exceed this total memory, it adds the query to a queue and considers the query with the largest predicted memory from the queue as the next query to submit.

$\bullet$ \textit{Qshuffler}: the Qshuffler scheduler~\cite{ahmad2011interaction}  uses a simple heuristic-based model to understand query interactions and schedule queries. Specifically, it first clusters all queries into $k$ types and assumes the queries from the same type have the same characteristics. Then, it represents the system state as a vector, counting the number of running queries of each type. At last, it scores all queued queries based on this vector and submits the one with the best score.

We did not compare with intrusive scheduling algorithms~\cite{kelley1959critical, ghodsi2011dominant, grandl2014multi, grandl2016graphene}, including RL-based learned methods~\cite{sabek2022lsched, mao2019learning, lyu2022fine}, because they require changes to the execution engine of the underlying DBMS that are impractical to implement in Postgres and Redshift. 

\sparagraph{Runtime Prediction Baselines}: We compared \modelname{} with the following baselines on concurrent query runtime prediction accuracy:

$\bullet$ \textit{Qshuffler}: the runtime predictor used by \textit{Qshuffler}~\cite{ahmad2011interaction}. 
%To compare our method with these baselines, we evaluate only the runtime prediction accuracy but not the end-to-end performance because they are not designed for query scheduling and cannot be trivially extended with our scheduling algorithm.

$\bullet$ \textit{GPredictor}: An ML-based state-of-the-art concurrent runtime predictor~\cite{zhou2020query}. Specifically, it represents the operators in the target query and its concurrently running queries as nodes and their interactions as edges. Then, it uses a graph neural network to propagate this graph's information and estimate each operator's runtime. 
We did not compare with other similar baselines~\cite{duggan2011performance, wu2013towards} because GPredictor has already demonstrated superior performance.

$\bullet$ \textit{Stage}: the staged runtime predictor~\cite{wu2024stage} mentioned in Section~\ref{subsec: featurization}. It only predicts the average runtime of a query without considering any concurrent query information. 

$\bullet$ \textit{Function}: expert-designed analytic functions~\cite{wu2013towards, duggan2014contender, ahmad2011predicting, ahmad2009query}, commonly used to predict runtime as a function of I/O, CPU, and memory usage. 
We combine the merits of existing approaches to derive a comprehensive analytic function. 
Specifically, this function jointly considers a wide range of query/system features (such as the estimated resource usage of queries, data sharing, and system resource capacities), which are independently modeled in prior works. 
It also contains several parameters to adjust for estimation errors and weight impacts on different resource types, optimized using multi-dimensional regression on the training data. The details can be found in Appendix~\cite{}. %\markos{maybe we should actually include the details here, at least for now, and decide whether/how much to cut when shortening the paper after fleshing out the intro}	

We tuned the hyperparameters of all baselines and our scheduler on a held-out validation trace. We ran experiments three times to evaluate performance and took the average performance.

%\smallskip
\sparagraph{Workloads:}
The most straightforward and fair approach to evaluate our scheduler and the baselines is to compare their end-to-end performance improvement over widely-used DBMSes on real workloads. 
Unfortunately, to the best of our knowledge, all existing approaches~\cite{gupta2009fair, ghodsi2011dominant, ahmad2011interaction, mehta2008bi} are evaluated on simulated workloads (e.g., having multiple clients issuing TPC-H/DS queries in a closed loop), which are not representative of real workload traces~\cite{van2024tpc, snowflake-nsdi20}.
Therefore, we conduct our experiments on the CAB~\cite{van2023cloud} and BRAD~\cite{yu2024blueprinting} workloads. 

$\bullet$ \textit{CAB}~\cite{van2023cloud} 
is a well-established cloud analytic benchmark for comparing data warehouse performance on OLAP query workloads.
It uses TPC-H benchmark~\cite{} to match the execution characteristics of Snowset~\cite{snowflake-nsdi20}, which contains real customer's query traces from Snowflake data warehouse but does not include the relevant tables or SQL statements.
Specifically, CAB contains a large pool of queries, generated from 22 TPC-H query templates.
CAB issues queries from the pool to match the typical query submission time patterns in Snowset traces.
Then, CAB uses different scales of TPC-H dataset to match the query execution time, CPU time, and data size of Snowset.
One limitation of CAB is that the TPC-H dataset is artificially generated and may not contain complex data distribution and correlations in real-world datasets. 
In addition, the 22 TPC-H query templates may not reflect the complexity and diversity of real-world queries. 
Therefore, we additionally conduct our experiments on the BRAD workload.

$\bullet$ \textit{BRAD}~\cite{yu2024blueprinting} proposed a realistic and diverse analytical workload using real-world data and queries. 
Specifically, BRAD scales the IMDB dataset~\cite{leis2015good} to $160GB$ in total size and generates a set of 1,000 OLAP queries with diverse query templates that resemble the IMDB JOB queries~\cite{leis2015good}.
Afterward, we generate query submission time based on the Snowset trace to derive a practical workload trace that closely matches Snowset's query submission times, completion times, and concurrency levels.

\begin{figure}[t]
	\centering
	\includegraphics[width=\columnwidth]{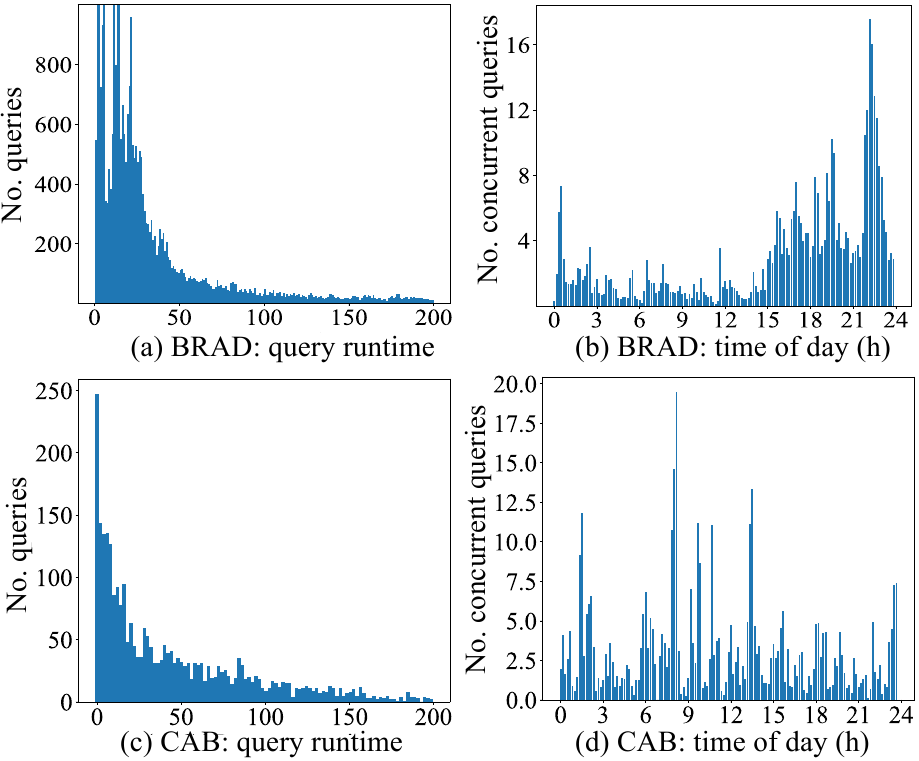}
    \vspace{-2em}
	\caption{Query runtime distribution and average number of concurrent queries (i.e., system concurrency level) throughout a day for BRAD and CAB workloads.}
	\label{fig:workload}
\end{figure}

\begin{figure*}[t]
	\centering
	\includegraphics[width=17.6cm]{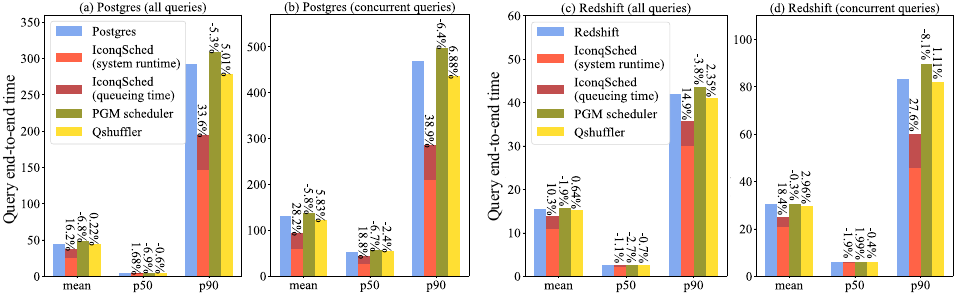}
	\caption{End-to-end performance of the non-intrusive schedulers in Postgres and Redshift for BRAD workload. The percentage improvements for the schedulers over the Postgres/Redshift are shown at the top of the bars.}
	\label{fig:overall_real}
\end{figure*}

\begin{figure*}[t]
	\centering
	\includegraphics[width=17.6cm]{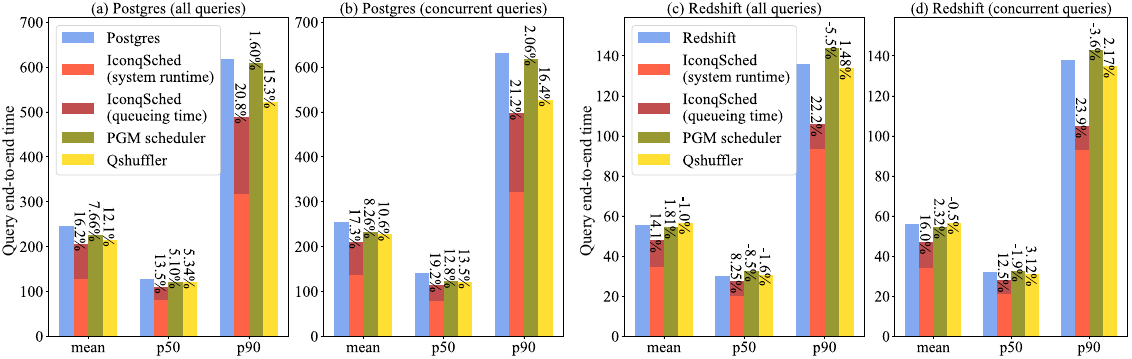}
	\caption{End-to-end performance of the non-intrusive schedulers in Postgres and Redshift for CAB workload. The percentage improvements for the schedulers over the Postgres/Redshift are shown at the top of the bars.}
	\label{fig:overall_real_cab}
\end{figure*}

Figure~\ref{fig:workload}-(a) shows the runtime distribution of a diverse set of queries in this workload, with runtimes ranging from a few seconds to a thousand seconds on Postgres. Roughly $40\%$  of the queries are short-running (i.e., < 10s), and $5\%$ of queries are long-running (> 200s) (not plotted in the figure for clarity).
The number of concurrent queries reflects the system load and directly impacts the scheduler's performance. Thus, we average the number of concurrent queries within a 5-minute interval and plot all such intervals throughout the last day in Figure~\ref{fig:workload}-(b). We can see that the system is ``underloaded'' (< 3 concurrent queries) for more than $50\%$ of time, but there exist spikes of more than 10 queries concurrently running in the system.
For both workloads, we obtained 7 days of execution trace, with an average of $5,000$ queries per day.
In experiments, we use the first 5 days of both traces as the training data to train our model and other baselines, the second last day as the validation trace to tune their hyperparameters, and the last day to evaluate their performance. 
In addition to these real-world traces, we conduct more adaptivity experiments on simulated workload traces, whose details are described in Section~\ref{subsec: exp-adaptivity}.

\subsection{End-to-end performance}
\label{subsec: exp-e2e}

To evaluate \sysname{} and the baseline systems, we first present their overall performance for executing the \textit{CAB} and \textit{BRAD} workloads in both Postgres and Redshift. Subsequently, we look at specific examples to understand how \sysname{} enhances performance compared to Postgres. For a fair assessment, we report the \fulltime{} (\textit{e2e-time}), which includes the scheduler's queueing time plus the \systime{} for a query.

\subsubsection{Performance on BRAD workload}
\label{sec:perf-brad} 
The overall performance on \textit{BRAD} workload in Postgres and Redshift is shown in Figure~\ref{fig:overall_real}. We report several metrics--average, median, and tail \textit{e2e-time}. 

\noindent \textbf{Performance in Postgres:}
Figure~\ref{fig:overall_real}-(a) shows the performance of all schedulers when executing the workload in Postgres. 
On average, \sysname{} achieves a $16.2\%$ improvement ($(Postgres - \sysname{}) / Postgres$) on all queries. Roughly $40\%$ of these queries are executed in isolation (i.e., without any other overlapping queries), so their \textit{e2e-time} cannot be improved by any non-intrusive scheduler.
%\markos{neither by any intrusive scheduler, perhaps?} \ziniuw{no, they can do fine-grained resource allocation in each operator and intra-query parallelization} \markos{makes sense}. 
After excluding these queries, \sysname{} achieves $28.2\%$ improvement on average on the remaining queries (labeled the \emph{concurrent queries}) in Figure~\ref{fig:overall_real}-(b).
Other baselines barely outperform Postgres because the complicated query interactions in the workload are hard to model with heuristics. 
On the p-50 metric of all queries, roughly all baselines achieve the same performance because more than $50\%$ of queries are short-running, which use a small amount of resources, and are not sensitive to the different concurrent states of the system. 
Thus, no scheduler is likely to improve the performance of these queries. Recent literature has made similar observations~\cite {ahmad2008modeling, ahmad2011interaction}. In contrast, when considering just long-running queries (those with p-90 runtime), \sysname{} can achieve a $33.6\%$ improvement over Postgres because of its accurate modeling of concurrent query performance and effective scheduling decisions. 
Further, \sysname{} can achieve $38.9\%$ improvement on the tail (p-90) for concurrent queries. 
It is worth noting that \sysname{} can improve the system runtime by $45\%$ on the tail, shown in the light orange bar of Figure~\ref{fig:overall_real}-(b). 
These results show that \sysname{} can effectively re-arrange query execution order to significantly speed up the overall workload execution. 
We provide detailed examples of how \sysname{} improves Postgres performance in Section~\ref{sec:casestudies}.

%Since the workload trace contains a large number of queries executed in isolation, which can not be improved by any non-intrusive scheduler, we provide the performance of queries with at least one concurrently running query ($60\%$ of the entire workload) in Figure~\ref{fig:overall_real}-(b) \markos{From Murali: explain what these concurrecnt queries are earlier, potentially have a ``waterfall'' of results}. For these queries, \sysname{} can achieve $28.2\%$ improvement on average, $18.8\%$ on the median, and $38.9\%$ on the tail. 
%It is worth noting that \sysname{} can improve the system runtime (excluding the scheduler's queueing time) by up to $45\%$. This suggests that \sysname{} can effectively re-arrange the execution order to significantly speed up the query execution. 
%\markos{Is this correct to claim? Don't the postgres numbers also include queueing time (whenever the MPL is exceeded)?}
    
\noindent \textbf{Performance on Redshift:}
We performed the same experiment on Redshift, as shown in Figures~\ref{fig:overall_real}-(c) and (d). Overall, \sysname{} achieves a $10.3\%$ average and $14.9\%$ tail improvement on all queries, and $18.4\%$ average and $27.6\%$ tail improvement on only the concurrently executed queries, whereas the other baselines do not improve over Redshift.
The performance improvement on Redshift is smaller than Postgres for two reasons. First, arguably, Redshift contains a state-of-the-art scheduler~\cite{wu2024stage, SaxenaRCLCCMKPN23, nathan2024intelligent}, leaving us less room for improving its performance with a non-intrusive scheduler. 
Second, Redshift queries generally execute faster on Redshift than on Postgres, and short-running queries are hard for \sysname{} to improve. This also explains why \sysname{} does not improve the median query \textit{e2e-time} in Figure~\ref{fig:overall_real}-(c) and (d).
%\srm{What about p90 runtime?  We focused on that for Postgres.}

\subsubsection{Performance on CAB workload}
\label{sec:perf-cab} 
The overall performance on \textit{CAB} workload is shown in Figure~\ref{fig:overall_real_cab}. 

\noindent \textbf{Performance in Postgres:}
Different from \textit{BRAD} workload, the \textit{CAB} workload only contains $22$ unique TPC-H query templates with much fewer joins and less complicated predicates than \textit{BRAD} queries. 
Furthermore, queries from the same template exhibit similar characteristics (e.g., scanning same tables, having same query plans, and using similar resource), making it easier to understand query interactions.
Thus, we observe that the non-intrusive scheduler baselines are able to improve the performance of Postgres, as shown in Figure~\ref{fig:overall_real_cab}-(a). 
In particular, \textit{Qshuffler} scheduler is able to cluster queries based on their corresponding query templates and capture interactions between query templates. \textit{Qshuffler} achieves up to $15.3\%$ improvement over Postgers. 
\sysname{} achieves $16.2\%$ average improvement, $13.5\%$ median improvement, and $20.8\%$ tail improvement over Postgres. 
\sysname{} still has a significant performance gain compared with the other baselines because of our accurate runtime predictor. 

When excluding the queries executed in isolation, we observe very close performance in Figure~\ref{fig:overall_real_cab}-(b) because \textit{CAB} is a heavier workload than  \textit{BRAD}, with only $5\%$ of the queries executed in isolation. This also explain why the overall query runtime of \textit{CAB} is much higher than that of \textit{BRAD} workload.

\noindent \textbf{Performance in Redshift:}
We observe similar results on \textit{CAB} workload. As shown in Figure~\ref{fig:overall_real_cab}-(c), \sysname{} achieves a $14.1\%$ average, $8.2\%$ median, and $22.2\%$ tail improvement over Redshift. 
This improvement is slightly higher than that on \textit{BRAD} workload because \textit{CAB} has a heavier query load and very few short-running queries.

\subsubsection{Example Scenarios}
\label{sec:casestudies} 

In the following, we provide two example scenarios derived from real executions of BRAD workload in Postgres to explain the performance gain of \sysname{}, demonstrating \sysname{}'s ability to avoid ``bad'' and seek ``good'' concurrent executions.

\begin{figure}[t]
	\centering
	\includegraphics[width=\columnwidth]{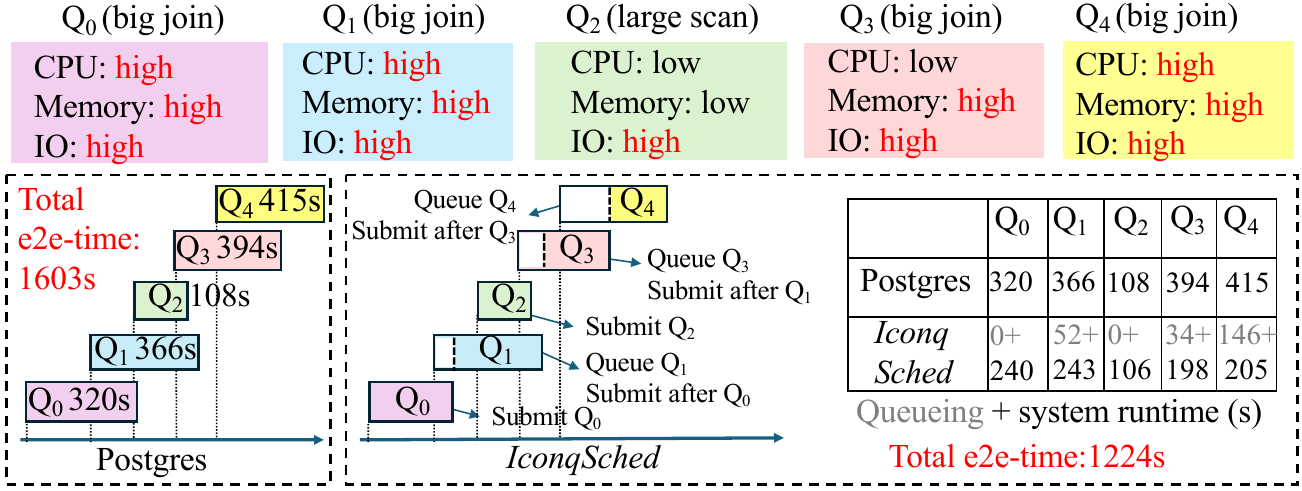}
	\caption{(Example 1) Delaying the execution of some queries can improve query performance.}
	\label{fig:cs_simple}
 \vspace{-2em}
\end{figure}

\textbf{First, sometimes executing queries sequentially can be \break more efficient than executing them concurrently.} 
For instance, executing two memory-intensive queries simultaneously can cause significant memory contention and even lead to disk spills, which greatly slows down the joins in both queries. Figure~\ref{fig:cs_simple} provides a concrete example involving five queries,  $Q_0, \ldots, Q_4$, executed within our workload. The legend at the top outlines the characteristics of each query and distinguishes them with different colors. The left panel shows their default execution in Postgres, where these queries are run concurrently, resulting in high resource contention and a long total \textit{e2e-time} of $1603s$. 

The right panel of Figure~\ref{fig:cs_simple} shows \sysname{}'s scheduling and execution decisions. \modelname{} accurately predicts the runtime of these queries both in sequential and concurrent executions,  identifying the benefits of delaying the execution of certain queries. As a result, when \sysname{} processes these queries, it decides to queue $Q_1$ and submit it after $Q_0$ is completed, queue $Q_3$ and submit it after $Q_1$, and queue $Q_4$ to be submitted after $Q_3$ finishes. This approach minimizes resource contention throughout the execution, thereby reducing the total \textit{e2e-time} to $1224s$. 

It's important to note that other scheduling methods, such as PGM
%\markos{wouldn't PGM identify the memory contention?} \ziniuw{not in the case where the system is not yet overloaded.}
and Qshuffler, lack the precise and detailed runtime prediction capabilities necessary to differentiate between sequential and concurrent execution of these queries.

\begin{figure}[t]
	\centering
	\includegraphics[width=\columnwidth]{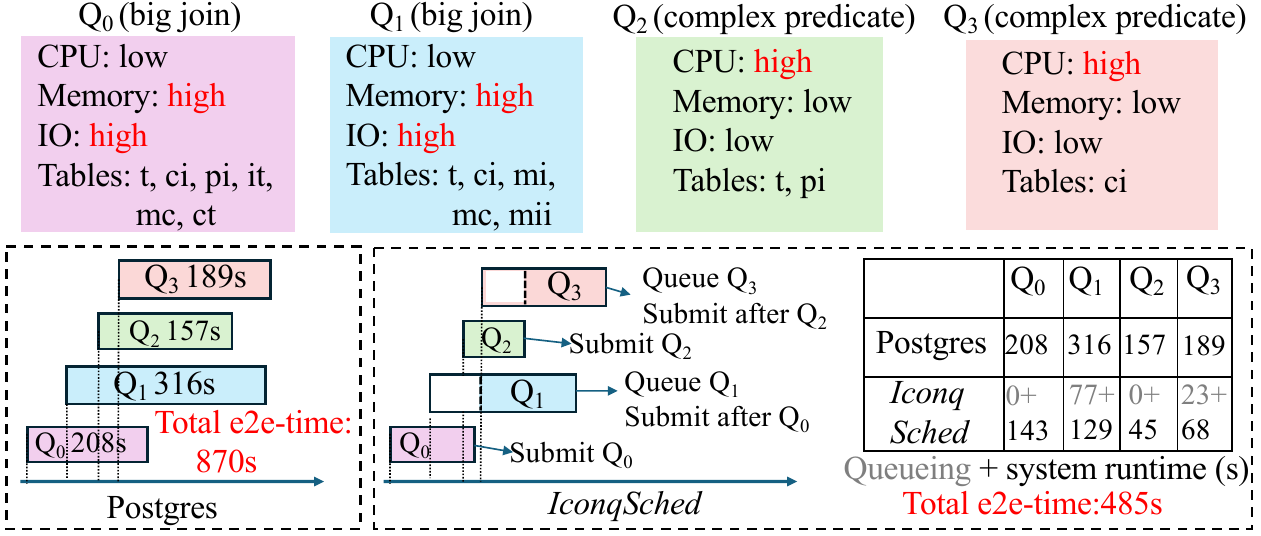}
	\caption{(Example 2) Optimizing resource usage and data sharing can improve query performance.}
	\label{fig:cs_resource}
 %\vspace{-2em}
\end{figure}

 \textbf{Second, optimizing different resource usage and
data sharing can improve query performance.}
An on-premises DBMS cluster contains a fixed number of different resource types (e.g., memory, CPU, I/O, bandwidth), and a query may use different amounts of each resource. For example, executing two queries with large joins may use up most of the memory, but the system may still efficiently process CPU-intensive queries (e.g., with complex filter predicates). 
As in previous example, Figure~\ref{fig:cs_resource} shows a example of four queries $Q_0, \ldots, Q_3$ in our workload executed on Postgres. $Q_0$ and $Q_1$ contain I/O- and memory-intensive big joins while $Q_2$ and $Q_3$ are CPU-intensive queries with complex filter predicates. 
Postgres executes $Q_0$, $Q_1$, $Q_2$, and $Q_3$ concurrently, leading to high resource contention and a long total \textit{e2e-time} of $870s$.

In contrast, \sysname{} can accurately model the interactions of these four queries and understand that executing $Q_0$ and $Q_1$ together is inefficient. 
Thus, \sysname{} decides to queue $Q_1$ until $Q_0$ finishes to avoid memory contention.
When $Q_2$ arrives, \sysname{} identifies the benefits of running $Q_2$ concurrently with $Q_0$, since they use different resources but access the same set of tables. Thus, \sysname{} decides to submit it directly. 
Similarly, \sysname{} understands that running $Q_2$ and $Q_3$ together is inefficient and decides to queue $Q_3$ and submit it after $Q_2$ finishes. 
Therefore, by using \modelname{}'s fine-grained concurrent runtime predictions, \sysname{} can implicitly optimize different types of resource usage and leverage data sharing. 
It is worth noting that instead of directly estimating each query's CPU or memory usage, \modelname{} implicitly captures these resource demands by accurately predicting the query runtime under different system states.

The above examples showcase the behavior of \sysname{} in specific scenarios. 
Unlike the above two examples, most cases do not have execution characteristics that are easy for humans to reason about. \sysname{} can still accurately judge which query in the waiting queue is optimal to submit and when to submit it.
The supplementary material provides a more complex scenario that \sysname{} frequently encounters.

\subsection{Adaptivity Analysis}
\label{subsec: exp-adaptivity}

\begin{figure}[t]
	\centering
	\includegraphics[width=8cm]{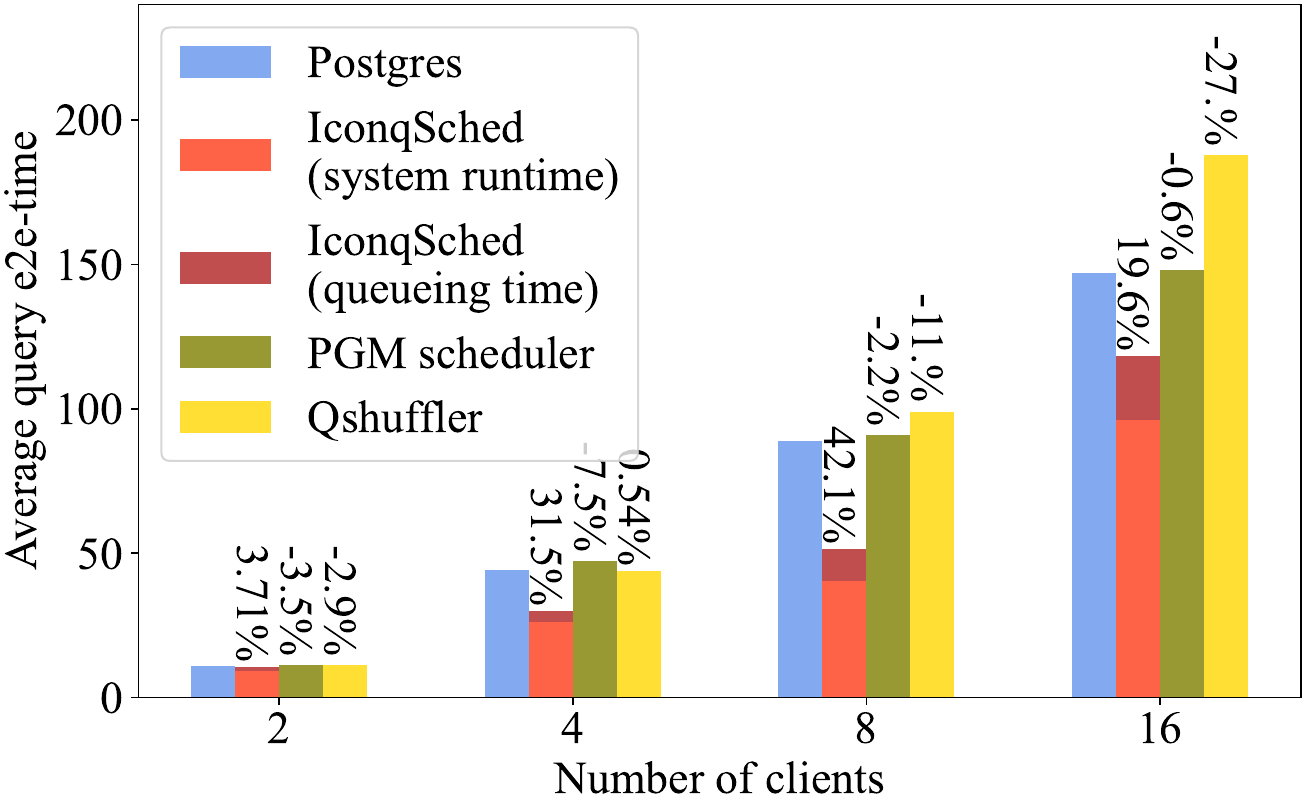}
	\caption{Performance of different schedulers with varied number of clients on executing BRAD queries in Postgres.}
	\label{fig:adaptivity_postgres}
 \vspace{-2em}
\end{figure}

To better understand the robustness of \sysname{} under various situations, we conducted adaptivity experiments with a varied number of clients. Our experimental setting is similar to many prior works~\cite{mehta2008bi, gupta2009fair} in that we simulate $k$ clients in parallel. 
Each client continuously issues queries in a closed loop to the same DBMS (Postgres or Redshift) instance. 
Specifically, each client randomly draws a query from the CAB/BRAD query pool, submits it to the DBMS instance, waits for it to finish, and immediately submits the next query. 
We chose the number of clients to be $k = 2, 4, 8, 16$, corresponding to a light load, an average load, a heavy load, and an extremely overloaded system state of the practical workload trace.
We ran each experiment for 10 hours.
Figure~\ref{fig:adaptivity_postgres} shows the average query runtime for different numbers of clients running BRAD queries on Postgres. 
Similar results on CAB queries and Redshift can be found in Appendix~\ref{sec:app-adaptivity}.
For $k=2$, \sysname{} performs similarly to Postgres because there are not enough queries in the waiting queue for \sysname{} to select a query to submit that positively interacts with the running queries. The slight improvement ($3.71\%$) is mainly due to situations similar to the example scenario of Figure~\ref{fig:cs_simple}.

\sysname{}'s performance improvement over Postgres increased significantly from two to eight clients because, with more clients, more queries are waiting in the queue, and our scheduler has a better opportunity to submit optimal queries at the optimal time. Moreover, with a median and heavy system load, cleverly managing system resource usage can lead to significant improvement in execution time. 
The performance improvement for 16 clients (19.6\%) is not as significant as for four or eight clients. Simulating with 16 concurrent clients severely overloaded the Postgres instance, with more than $10\%$ of the queries timing out (> 1000s) in the original Postgres execution. Because of the extremely heavy load, \sysname{} frequently queues queries to offload the instance. As a result, many queries are queued for a long time and have to be submitted due to the starvation penalty (Equation~\ref{equ: score}). These queries are not necessarily the optimal queries to submit at the optimal time. Therefore, we observe a decrease in performance gain for \sysname{} in the case of 16 clients. We plan to explore how to address this issue in future work. We did not experiment with more than 16 clients because $k=16$ already clearly exceeded the capacity of the Postgres instance.

For the Qshuffler and the PGM scheduler, it is difficult to understand the complex interaction of heavy system load using a heuristic-based runtime predictor or a simple mechanism. 
Instead, they can sometimes increase the queueing time.
As a result, these two baselines frequently make suboptimal decisions and sometimes perform worse than Postgres.
%We observe similar behavior on Redshift, which can be found in the supplementary material. 

\subsection{Performance of Runtime Predictors}
\label{subsec: exp-accuracy}

In this section, we evaluate \modelname{} against the baselines mentioned in Section~\ref{subsec: exp-setup}.
We first report the overall performance on the BRAD workload executed in Postgres and then evaluate the generalizability of each method. 
We observe similar performance trends on the CAB workload and Redshift.
%Due to space limitations, we put similar results on the CAB workload and Redshift in our technical report~\cite{}.

\smallskip
\noindent \textbf{Overall performance:}
Following the convention of prior works~\cite{wu2024stage, negi2023robust, wu2023factorjoin, wu2020bayescard, han2021cardinality}, we judge the prediction accuracy of each method at p50, p90, and p95, using two well-recognized metrics: i) \textit{absolute-error}, which measures the absolute difference between predicted runtime and true runtime: $|pred - true|$; and ii) \textit{Q-error}, which measures their relative difference: $max\{pred/true, true/pred\}$. On both metrics lower is better, with $0$ and $1$ being optimal, respectively.

\begin{figure}[t]
	\centering
	\includegraphics[width=\columnwidth]{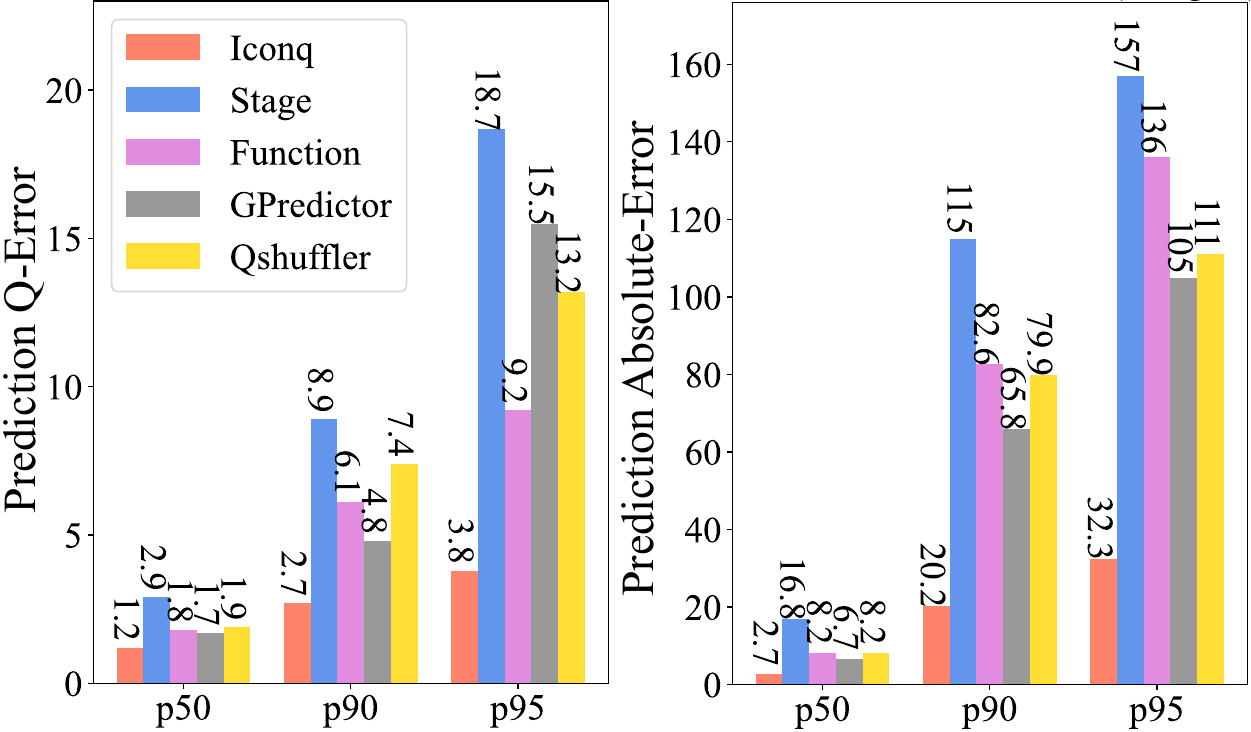}
	\caption{Runtime predictors' performance on concurrent queries of BRAD workload in Postgres.}
	\label{fig:acc_postgres}
\end{figure}

Figure~\ref{fig:acc_postgres} shows the performance of \modelname{} and the baselines on all concurrent queries of the testing workload ($\sim5000$ queries). \modelname{} achieves significantly better performance on all metrics. We observe a $1.5x - 2.4x$ and $2.4x-4.9x$  difference on \textit{Q-error} for median and tail (p-90), respectively. Similarly, on \textit{absolute-error}, we observe a $2.5x - 6.2x$ improvement on the median and $3.3x-4.9x$ improvement on the tail. 
\textit{Stage} performs the worst because it only estimates the average query runtime without considering concurrent query information.
We also compared other aspects of runtime predictors, such as the training time, model size, and inference speed in Appendix~\ref{sec:app-runtime}.

\smallskip
\noindent \textbf{Robustness and generalizability:}
In addition, we conducted two generalizability experiments to demonstrate the robustness of \modelname{}.

\begin{figure}[t]
	\centering
	\includegraphics[width=8cm]{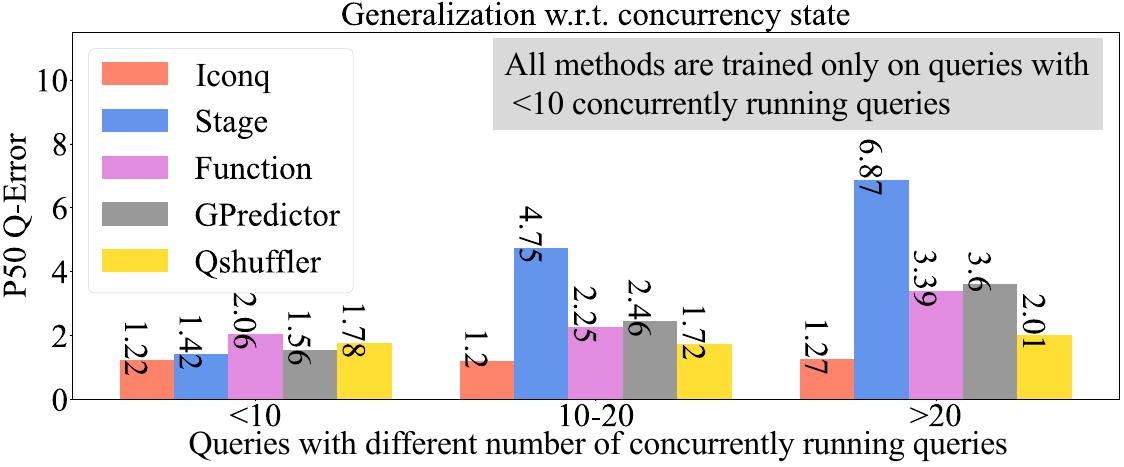}
	\caption{Runtime predictors' generalizability to queries executed under heavier concurrency state.}
	\label{fig:gen_concur}
  %\vspace{-1em}
\end{figure}

First, we evaluate these runtime predictors' generalizability to heavier concurrency states, i.e., queries with more concurrently running queries.
Specifically, we train these predictors on queries with $<10$ concurrently running queries and test on queries with $10-20$ and $>20$ concurrently running queries.
Figure~\ref{fig:gen_concur} shows that \modelname{} remains very accurate even for unseen heavier concurrency states. This is because \modelname{} uses the hidden state of a bi-directional LSTM to represent concurrent and system state changes that can better extrapolate to an unseen concurrency state.
We provide more detailed reasons for \modelname{}'s robustness in Section~\ref{subsec: lstm}.
We also find \textit{Qshuffler}'s performance relatively stable because its heuristics are robust against changing concurrent states. However, other baselines experience a significant decrease in accuracy when tested on heavier concurrent states.

\begin{figure}[t]
	\centering
	\includegraphics[width=8cm]{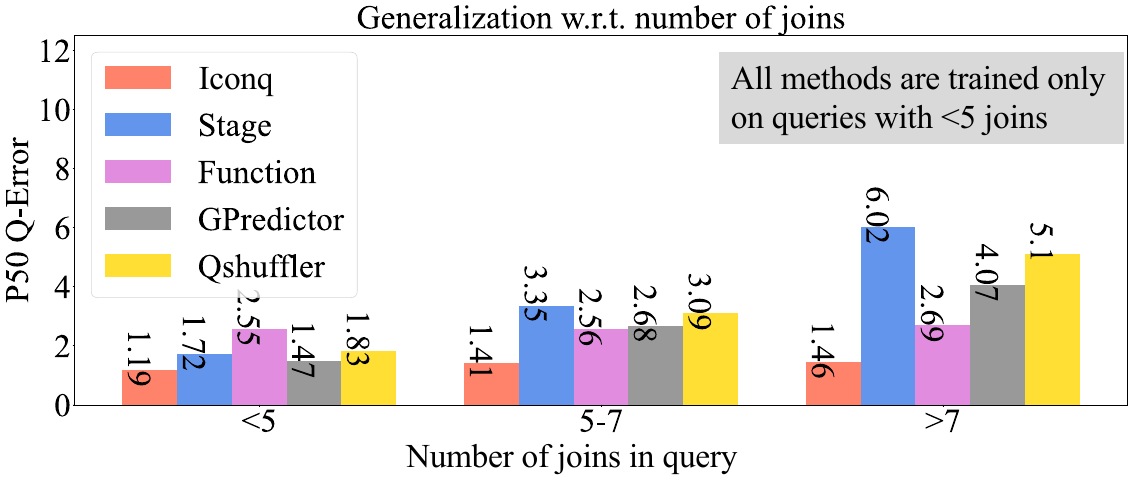}
	\caption{Runtime predictors' generalizability to queries with more complex templates.}
	\label{fig:gen_template}
 %\vspace{-1em}
\end{figure}

Second, we evaluate these methods' generalizability to more complex query templates (e.g., more table joins). We train them on queries with $<5$ table joins and test on queries with $5-7$ and $>7$ table joins.
Figure~\ref{fig:gen_concur} shows that \modelname{} remains very accurate even for harder query templates because the model feature captures the inherent properties of the queries (e.g., average runtime, cardinalities, and plan structure) that can be better extrapolated to unseen query templates.
We also find \textit{Function}'s performance relatively stable because this simple method does not encode any information about the query template. 
However, other baselines experience a significant decrease in accuracy when tested on harder templates.
\textit{GPredictor} featurizes the query plan as a graph. For queries with more joins, their corresponding graphs get much larger. 
We hypothesize that the trained GCN model may overfit small graphs and extrapolate poorly to larger ones.
\textit{Qshuffler} uses a simple clustering model to group similar queries into clusters. For more complex queries, the trained model cannot easily cluster them and will lead to very poor performance.

%\markos{In general, all of the descriptions in the evaluation feel a bit wordy. Once we have a better idea of the overall length (after adding intro, conclusion etc) I can take a shortening pass if you want.}

%% file: background.tex
\section{Related work}
\label{sec:background}

To highlight this work's novelty, we review the related work in runtime prediction and scheduling for OLAP queries. 

\noindent \textbf{Query runtime prediction:} 
Much of the prior work focuses on predicting the runtime of a single query without considering its concurrent state. Traditional methods typically rely on manually-derived heuristics and statistical models to analyze relational operators~\cite{DugganPCU14, WuCZTHN13, AkdereCRUZ12, LiKNC12}. On the other hand, machine learning models can predict the runtime of a query with greater accuracy than traditional approaches, but they experience high inference latency~\cite{MarcusP19, SunL19, neo, bao, balsa, HilprechtB22, zhao2022queryformer, wu2024stage}.%Recently, \textit{Stage}~\cite{wu2024stage} adopts a hierarchy of different models to predict query runtime with high accuracy and low latency. It demonstrates its effectiveness and practicality inside Redshift. 

Predicting the runtime of a target query given its concurrent state (i.e., the system state and the concurrently executed queries) is a much harder problem than single query prediction. Most of the existing methods are based on simple heuristics. Analytic functions/models~\cite{wu2013towards, duggan2014contender, ahmad2011predicting, ahmad2009query} are commonly used to predict concurrent runtime as a function of I/O, CPU, and memory usage. BAL~\cite{duggan2011performance} uses buffer access latency to capture the effect of disk I/O. Qshuffler~\cite{ahmad2011interaction} clusters queries into different types and approximates query interactions between these types. Such simple approaches cannot accurately predict the runtime of queries in practical workloads with complex query interactions and system state changes. A recent work~\cite{zhou2020query} proposed using a graph neural network (GNN) to understand complex interactions among concurrently executing queries. It has been shown to be significantly more accurate than the traditional methods. However, this GNN-based approach is not well suited to our work. First, it assumes perfect knowledge of queries that arrive after the target query, which is impractical in our setting. Second, it cannot understand system state changes at different timestamps, which can produce inaccurate predictions. 
%Third, the GNN model predicts the runtime of each operator in a query's execution plan, whereas we need the overall runtime of the entire query \srm{if you know the runtime of each operator, isn't it trivial to get the runtime of the plan?}.%To address these drawbacks, \sysname{} designs a novel concurrent query featurization method and adapts an LSTM model to understand query interactions and system state changes.

\smallskip
\noindent \textbf{Query scheduling:}
%We  roughly categorize  existing query schedulers as either intrusive or non-intrusive. 
%Intrusive schedulers require access or changes to the execution engine, which must be implemented inside the DBMS. 
Traditional intrusive schedulers generally use simple heuristics or analytical models to estimate the needs of each query/operator and allocate appropriate resources~\cite{kelley1959critical, ghodsi2011dominant, grandl2014multi, grandl2016graphene, SaxenaRCLCCMKPN23}. 
Recent approaches~\cite{sabek2022lsched, mao2019learning, lyu2022fine} propose representing query plans as graphs and using deep reinforcement learning (RL) to make fine-grained scheduling decisions.
They can cleverly allocate appropriate resources to different query operators, which significantly improves over traditional methods.
However, these RL-based methods require extensive training and exploration inside the DBMS instance before achieving satisfactory performance, which may not be affordable for many users. 
Moreover, as mentioned in Section~\ref{sec:intro}, intrusive schedulers come with a high engineering cost.
%Moreover, system engineers require months to implement and verify these intrusive algorithms before they are available to end users, and new implementations need to be done for every system.

%In contrast, non-intrusive schedulers do not need access to internal DBMS statistics or require modifications to the execution engine. 

Many non-intrusive approaches also use simple rules or heuristics to schedule queries and control query admission, such as first-in-first-out, shortest-query-first~\cite{SaxenaRCLCCMKPN23}, and fair scheduling~\cite{gupta2009fair, ghodsi2011dominant}. Others use analytic models and regression models to estimate query cost/runtime and resource consumption in order to control the multi-programming level (MPL) and perform query admission~\cite{denning1980working, denning1976optimal, mehta2008bi, SQLserver2017, zhang2012discovering, haritsa1993value, chi2013distribution, peha1991scheduling}. QShuffler~\cite{ahmad2008modeling, ahmad2011interaction} uses heuristics to approximate query interactions and leverages them to schedule queries that can minimize the overall runtime. These heuristic-based non-intrusive scheduling algorithms 
%are largely based on simple heuristics and rules that 
can easily make sub-optimal decisions. In contrast, \sysname{} leverages its fine-grained runtime predictions to accurately understand query performance under different concurrent query loads to decide which queries to execute and when to execute them. %\markos{Again it may be good to have a sentence at the end that concisely underlines what we do different than related work}
%\markos{isn't our short query acceleration a heuristic?}
%\ziniuw{true, let's just say the cost model}

%\subsection{LSTM models}
%\label{subsec:background-lstm}

%% file: conclusion.tex
\vspace{0.5em}
\section{Conclusion}
\label{sec:conclusion}

\smallskip

This work introduces \sysname{}, a  non-intrusive scheduler designed to enhance the performance of a DBMS by effectively rearranging the execution order and submission time of queries. \sysname{} employs a novel fine-grained predictor, \modelname{}, which treats the DBMS instance as a black box and  predicts the \systime{} of concurrently executed queries on that instance. Experiments demonstrate that \sysname{} can significantly improve the performance of Postgres and Redshift on a realistic OLAP workload. Although not a focus of this work, we believe \modelname{} has significant additional benefits in other DBMS tasks, such as query optimization and maintaining service level objectives (SLOs).

%% file: appendix.tex
\section{Details of scheduling algorithm}
\label{sec:app-algo}
In this section, we will illustrate the procedure for using the aforementioned runtime predictor to decide which queries in the waiting queue to submit for execution and when to submit them. We first explain the problem formulation for scheduling an online stream of queries and its technical challenges. Then, we provide the details of \sysname{}'s scheduling algorithm. 

\subsection{Problem formulation}

Finding the optimal solution is far from trivial.
For example, a brute-force solution would be exhaustively enumerating the entire search space of all possible decisions, predicting the runtime of each corresponding situation using our predictor, and selecting the best decision with minimal overall predicted runtime. 
However, this approach is impractical for several reasons. 
First, there are $n \choose k$ ways to select $k$ queries to submit out of a queue with $n$ queries, which is unaffordable to enumerate. Second, the decision space for when to submit them is infinite because any future timestamp can be a candidate choice. Therefore, it is impractical to enumerate the entire search space of candidate decisions.
In addition to the exploding search space, the scheduler faces another challenge --- uncertainty in the future. 
Since we consider the problem of ingesting a stream of queries online, the optimal decision at the moment might be sub-optimal if particular queries come in the future. 
Thus, our scheduler needs to make robust decisions given the imperfect information at the current timestamp.
It is worth noticing that predicting queries arriving in the future is an orthogonal field of research~\cite{ma2018query,huang2024sibyl}.
In this work, we will not make any assumptions about future queries. We will explore integrating a workload forecaster with our scheduler in future work.

To efficiently address the challenges, we design a new scheduling algorithm with linear time complexity to efficiently use our fine-grained runtime predictor and effectively approximate the optimal solution.
It is worth noticing that the existing non-intrusive scheduling algorithms~\cite{SaxenaRCLCCMKPN23, gupta2009fair, mehta2008bi, denning1976optimal, ahmad2011interaction} are generally based on expert-designed rules and heuristics not only because of the exploding search space but also because they do not have an accurate runtime predictor to backup a more principled algorithm design.

\subsection{Algorithm details}

\begin{algorithm}[h]
	\caption{Greedy scheduling algorithm}
	\scriptsize
	\label{alg:scheduler}
	\begin{algorithmic}[1]
		\Require $\mathbf{Stage}$:  the trained \textit{stage} model to predict the average runtime of a single query
		\\$\mathbf{LSTM}$:  the trained \textit{LSTM} model to predict the runtime of a query with its concurrent queries
		\\ $\tau_{short}$: threshold for short running queries
		\\ $\lambda$: penalty for queueing a query too long
		\\ $lookahead$: number of future timestamps to compare with
		\State At timestamp $t$, there are $n$ queries running in the system $RQ$ = \{$RQ_1, \ldots, RQ_n$\} and $m$ queries waiting in the queue $WQ$ = \{$WQ_1, \ldots, WQ_m$\}
		\State Let $QT(WQ_i)$ denote the queueing time of $WQ_i$
		\State Let $\mathbf{Stage}(WQ_i)$ denote the predicted average runtime of $WQ_i$
		\State Let $\mathbf{LSTM}(WQ_i, RQ, t)$ denote the predicted execution time of $WQ_i$ given $RQ$ queries concurrently running in the instance at timestamp $t$
		\\
		\Function{Score}{$RQ$, $WQ_i$, $t$}
		\State $\Delta_1 \gets \mathbf{LSTM}(WQ_i, RQ, t) - \mathbf{Stage}(WQ_i)$
		\LComment{$\Delta_1$ measures the benefits on $WQ_i$ of submitting it now v.s. the average case}
		\State $\Delta_2 \gets \sum_{j} \mathbf{LSTM}(RQ_j,  \{WQ_i\} \bigcup RQ / \{RQ_j\}, t) - \mathbf{LSTM}(RQ_j, RQ / \{RQ_j\}, t)$ 
		\LComment{$\Delta_2$ measures how much impact will $WQ_i$ have on each $RQ$}
		\State $s \gets \Delta_1 + \Delta_2 - \lambda * QT(WQ_i)$
		\LComment{Score considers $\Delta_1$, $\Delta_2$ and a penalty for queueing a query too long}
		\State \Output $s$
		\EndFunction
		\\
		\Function{Should\_submit}{$RQ$, $WQ_i$, $t$}
		\LComment{Decide whether it is better to submit $WQ_i$ now  (at timestamp $t$) or later.}
		\For{$j = 1$; $j <= lookahead$; j++}
		\State $t_j  \gets$ predicted finishing time of next $j$-th finishing query in $RQ$
		\State $\delta_1 \gets \mathbf{LSTM}(WQ_i, RQ, t) -  \mathbf{LSTM}(WQ_i, RQ, t_j) - (t_j - t)$
		\LComment{$\delta_1$ measures the benefits on $WQ_i$ of submitting at $t$ v.s. at $t_j$}
			\State $\delta_2 \gets \sum_{k} \mathbf{LSTM}(RQ_k,  \{WQ_k\} \bigcup RQ / \{RQ_k\}, t) - \mathbf{LSTM}(RQ_k,  \{WQ_k\} \bigcup RQ / \{RQ_k\}, t_j)$ 
		\LComment{$\delta_2$ measures how much impact on each $RQ$ when submitting $WQ_i$ at $t$ v.s. at $t_j$}
		\If{$\delta_1 + \delta_2 > 0$}
		\LComment{Positive value suggets submitting at future timestamp $t_j$ is better than $t$}
		\State \Output False
		\EndIf
		\EndFor
		\State \Output True
		\EndFunction
		\\
		
		\Function{Ingest}{$RQ$, $WQ$, $t$, $\tau_{short}$}
		\State $S \gets $ [ ]
		\State Use $\mathbf{LSTM}$ to predict the runtime of all $RQ_j$: $RT(RQ_j)$. 
		\ForAll{$WQ_i \in WQ$}
		\If{$\mathbf{Stage}(WQ_i) < \tau_{short}$}
			\LComment{Directly submit short-running queries.}
			\State \Output $WQ_i$
		\EndIf
		\If{\textsc{Should\_submit}($RQ$, $WQ_i$, $t$)}
			\State $s \gets$ \textsc{Score}($RQ$, $WQ_i$, $t$)
			\State $S \gets S$ add $s$
		\EndIf
		\EndFor{}
	
		\If{S == []}
		\LComment{No query in the queue should be submitted now}
		\State \Output None
		\Else{}
		\LComment{Submit the query with optimal (smallest) score}
		\State $selected \gets argmin(S)$
		\State \Output $WQ_{selected}$
		\EndIf
		\EndFunction
		
		\\ 
		\While{True}
		\LComment{Continuously ingesting an online stream of queries.}
		\State $need\_ingest \gets False$
		\If{A query $Q_i$ arrives at time $t$}
		\State $WQ \gets WQ$ add $Q_i$
		\State $need\_ingest \gets True$
		\EndIf
		\If{A query $Q_i$ finishes at time $t$}
		\State $RQ \gets RQ$ remove $Q_i$
		\State $need\_ingest \gets True$
		\EndIf
		\If{$need\_ingest$}
		\State $Q_{selected} \gets$ \textsc{Ingest}($RQ$, $WQ$, $t$, $\tau_{short}$)
		\While{$Q_{selected} \neq$ None}
		\State Submit $Q_{selected}$ to the system
		\State $RQ \gets RQ$ add $Q_{selected}$
		\State $WQ \gets WQ$ remove $Q_{selected}$
		\State $Q_{selected} \gets$ \textsc{Ingest}($RQ$, $WQ$, $t$, $\tau_{short}$)
		\EndWhile
		\EndIf
		\EndWhile
	\end{algorithmic}
\end{algorithm}

Recall that the goal of our scheduler is to decide which queries in the waiting queue to submit for execution and when to submit them.
Specifically, given n queries running in the system \{$RQ_1, \ldots, RQ_n$\} and m queries waiting in the queue \{$WQ_1, \ldots, WQ_m$\}, the scheduler needs to decide which queries (if any) to submit to the system. 
Because the search space for candidate queries to submit is exponential, and the search space for when to submit them is infinite, we design an efficient and effective greedy scheduling algorithm. 
Our algorithm submits only one optimal query at a time to avoid the exponential search space. After submitting the optimal query for execution, our scheduler immediately reruns the algorithm to decide on the second optimal query (if any).
In addition, to avoid the infinite search space of when to submit a query, our scheduler only considers whether to submit a query whenever it ingests a new query or an old query finishes.
We will describe the pros and cons of this design after providing a detailed algorithm description.

\smallskip
\noindent \textbf{Candidate queries for submission:}
In the online ingestion phase at the current timestamp $t$, for each query $WQ_i$ in the waiting queue $WQ$, our scheduling function needs to decide whether to submit $WQ_i$ for execution or keep queueing it.
%We determine whether $WQ_i$ is beneficial to submit now by comparing the overall changes in the runtime of all queries running in the system. 
We first define a tunable hyperparameter $lookahead$ representing the number of future timestamps to compare with (line 16). 
Recall that our scheduler does not assume any knowledge about future queries and their arrival rate. 
Thus, if the scheduler does not submit $WQ_i$ now, it will reconsider $WQ_i$'s submission only when one of the running queries finishes. 
We denote the predicted finishing timestamps as $t_1, \ldots, t_{lookahead}$, where $t_j$ corresponds to the next $j$-th finishing query.
Thus, the algorithm needs to compare the benefit of submitting $WQ_i$ now at $t$ or in the future at $t_j$. 

Specifically, for each future timestamp $t_j$, we calculate the difference in the predicted runtime of $WQ_i$ between submitting at timestamp $t$ and $t_j$, denoting as $\delta_1$ in line 17. Then, we calculate the impact of submitting $WQ_i$ at $t$ and $t_j$ on each running query $RQ_k$ as $\mathbf{LSTM}(RQ_k, \{WQ_k\} \bigcup RQ / \{RQ_k\}, t_j)$. Next, we aggregate the impact difference on all running queries as $\delta_2$ in line 20.
Thereafter, $\delta_1 + \delta_2 > 0$ for timestamp $t_j$ suggests that submitting $WQ_i$ at $t_j$ is more beneficial than submitting now at $t$ and thus, we should keep $WQ_i$ in the queue (lines 22-24). 
Otherwise, $\delta_1 + \delta_2 \leq 0$ for all future timestamps $t_1, \ldots, t_{lookahead}$ would suggest that it is more beneficial to submit $WQ_i$ now (line 25). In this case, we will consider $WQ_i$ as a candidate query for submission (lines 34-36). 

Hyperparameter $lookahead$ balances the algorithm's efficiency and accuracy. A higher $lookahead$ will consider more timestamps in the future and better justify the decision to submit a queued query. However, a higher $lookahead$ value will impose a larger overhead for the scheduling algorithm. 
In our experiments, we set $lookahead = 2$ and empirically verified that a larger  $lookahead$ does not significantly improve the scheduler's performance.

\smallskip
\noindent \textbf{Scoring function:}
After deriving a set of candidate queries for submission, our algorithm needs to rank them based on some scoring function and select the optimal one to submit for execution. 
Since the objective of our scheduler is to minimize the average/sum runtime of all queries, we define our scoring function as follows. 
For each candidate query $WQ_i$, we first measure the impact that the current system state will have on $WQ_i$ as the difference between the predicted concurrent runtime and the average runtime of $WQ_i$ ($\Delta_1$ in line 6). 
A small negative value of $\Delta_1$ suggests that the current system state has a positive impact on $WQ_i$ that makes its runtime shorter than average.
Then, we measure the impact that $WQ_i$ has on the current system state as the total changes in the runtime of running queries $RQ$ ($\Delta_2$ in line 8).
Thus, $\Delta_1 + \Delta_2$ corresponds to the benefit in overall query runtime if we submit $WQ_i$ now.
In addition, to prevent a query from starving in the waiting queue for too long, we introduce a hyperparameter $\lambda$ that penalizes starving queries.
In this case, the candidate query with the smallest score will be the optimal query to reduce the average/sum runtime.
It is possible to modify this scoring function to optimize the median or tail runtime of all queries~\cite{koenker2005quantile, ben2009robust}. We leave this exploration as a future work.

\smallskip
\noindent \textbf{Query submission for execution:}
Two cases exist in which the scheduler will submit a query for execution, which is determined by the \textsc{Ingest} function (lines 27-43).
First, our scheduler directly submits short-running queries (lines 31-33). 
Our scheduling algorithm needs to invoke the LSTM model to make decisions, incurring an additional $10-100ms$ overhead. This overhead may be negligible for long-running queries but can be significant for short-running queries (e.g., < 5s).  In addition, we observe that the runtime of shorting-running queries is very stable and does not change much under different system loads and concurrent states because they use very limited resources. 
Prior works make the similar observations~\cite{ahmad2008modeling, ahmad2011interaction}.
Therefore, the scheduler will have a limited impact on shorting-running queries, so we will directly submit them.

Second, for long-running queries, we follow the aforementioned procedure to derive a subset of candidate queries, rank them based on the scoring function, and submit the optimal queries with small scores.
It is worth noticing that our scheduler submits one query at a time to avoid the exponential search space. 
Thus, after submitting the optimal query, we \textit{immediately and recursively} invoke the \textsc{Ingest} function to determine the next most optimal query (lines 56-60).
The recursive calls stop until there is no candidate query, meaning all queries in the waiting queue $WQ$ are more beneficial to submit at later timestamps.

\smallskip
\noindent \textbf{Candidate timestamps to submit queued queries:}
Our scheduler invokes our \textsc{Ingest} function to decide whether to submit a query whenever it ingests a new query or an old query finishes.
This design considers both efficiency and effectiveness.

Since there are infinitely many possible timestamps for submitting queries, we need to discretize the time interval for efficiency.
A naive approach will divide the timestamps with equal intervals (i.e., invoking  \textsc{Ingest} every 5 seconds). 
However, this approach has several downsides.
First, this may lead to non-trivial queueing time, especially for short-running queries. Thus, we need to invoke our \textsc{Ingest} function immediately whenever we ingest a new query so that we can decide whether to directly submit it or not.
Second, this approach has redundant \textsc{Ingest} invocations, increasing overhead. For example, if there is one long-running query $RQ_1$ that may incur a similar concurrent state throughout its execution. Calling  \textsc{Ingest} function multiple times during $RQ_1$ execution is redundant because the scheduling algorithm is very likely to make similar predictions and make similar decisions. 
Thus, we only need to invoke \textsc{Ingest} when there is a change in concurrent state (i.e. when a query finishes execution).

\smallskip
\noindent \textbf{Analysis:}
Our greedy scheduling algorithm is efficient and effective at the same time. 
Compared to the exhaustive search with exponential time complexity, our algorithm only invokes the $\mathbf{LSTM}$ model $|RQ| + lookahead * |WQ|$ times in a batch. In the experiment, we set $lookahead$ to $2$, making the overall time complexity linear with respect to the total number of queries. 
Compared to the heuristic-based schedulers~\cite{SaxenaRCLCCMKPN23, gupta2009fair, mehta2008bi, ahmad2011interaction}, our scheduler presents a more principled framework. It relies on fine-grained runtime predictors and makes accurate scheduling decisions accordingly.
In addition, our scheduling algorithm is also very explainable, easy to debug, and can support incrementally adding user-defined rules.

\section{Additional details on experiments}

\subsection{Replaying workload trace}
\label{sec:app-replay}

In this section, we describe how to replay the Snowset trace on IMDB data and queries.
Our approach can be generally applied to arbitrary traces with query start/finish time.

\smallskip
\noindent \textbf{Problem formulation:}
Big companies release their workload traces from their customers~\cite{van2024tpc, snowflake-nsdi20}, which closely represent the industrial needs for DBMS. However, due to data privacy, these traces only contain a log of query start/finish time on each DBMS instance without providing the SQL statement and data tables.
Therefore, the goal is to use the open-source data and queries to replay this workload on our selected DBMS instance so that every query has the same start time and a very close finish time, i.e., matching the query runtime and concurrency level as close as possible. 
In this paper, we take a representative trace (DB instance id: 1453912639619907921) from Snowset~\cite{snowflake-nsdi20}, which logs the submission and finish time for all queries in a cluster over 7 days. We use the scaled IMDB dataset (100GB) and a pool of 1k diverse OLAP queries from BRAD~\cite{yu2024blueprinting} to replay this workload trace.

\smallskip
\noindent \textbf{Technical challenges:}
Matching the query runtime of a workload is not trivial because of the complex behavior of executing queries concurrently. 
For example, to match a query with $10s$ runtime on Snowset, a naive approach would be to draw a query $Q$ from our IMDB query pool with an average runtime of $10s$. However, due to other concurrently executed queries, $Q$ may execute for $30s$. 
Therefore, we must draw a query with $10s$ runtime when concurrently executed with other queries.

\smallskip
\noindent \textbf{Our approach:}
We leverage our trained concurrent runtime predictor \modelname{} to draw a query from our IMDB query pool that most closely matches the runtime of the target query in Snowset when executed concurrently with other queries. 
Specifically, we first collect training data of \modelname{} and \textit{Stage} model by randomly spawning $2, 4, 8, 16$ clients that issue queries to the DBMS instance.
After training the \modelname{} and \textit{Stage} model, we use an iterative approach to match the target Snowset workload.
In the first iteration, we use \textit{Stage} to draw queries from IMDB query pool with average runtime matching Snowset queries.
Then, we use \modelname{} to predict the runtime of each query given its concurrently executed queries. 
We check for all queries whether their predicted runtime matches their corresponding Snowset queries (here, we say it is a match if the predicted runtime is within $20\%$ of the target runtime).
For every query whose predicted runtime is smaller than the target Snowset query, we draw a query with a larger runtime and vice versa.
In the following iteration, use \modelname{} to predict the runtime of the adjusted queries and make further adjustments if necessary.
We stop the iteration upon convergence, where most of the queries' runtime match their corresponding Snowset queries. 
We set a maximum iteration threshold to stop the process after $100$ iterations.

\begin{figure}[t]
	\centering
	\includegraphics[width=\columnwidth]{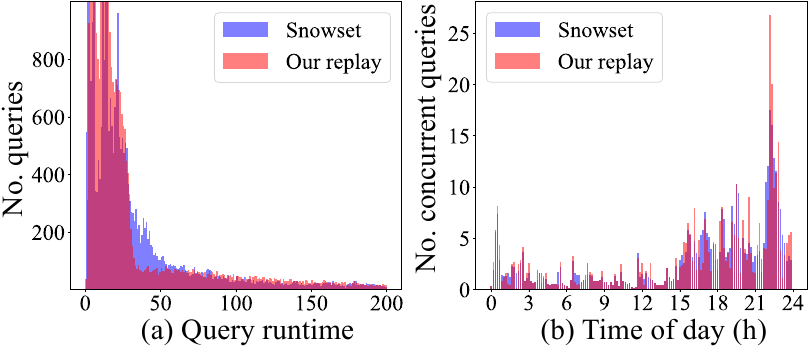}
    \vspace{-1em}
	\caption{Query runtime and the average number of concurrent queries throughout a day.}
	\label{fig:workload_appendix}
\end{figure}

\smallskip
\noindent \textbf{Results:}
Figure~\ref{fig:workload_appendix}-(a) shows the runtime distribution of a diverse set of
queries in this workload, ranging from a few seconds to a thousand seconds on Postgres. 
We can see that the overall query runtime distribution of our replayed workload on IMDB data and queries roughly matches that of the original Snowset workload.
The number of concurrent queries reflects
the system load and directly impacts the scheduler’s performance.
Thus, we average the number of concurrent queries within a 5-minute interval and plot all such intervals throughout the last day in Figure~\ref{fig:workload_appendix}-(b).
We can see that the concurrent query distribution of our replayed workload on IMDB data and queries roughly matches that of the original Snowset workload.

\subsection{Runtime prediction using analytical functions}
\label{sec:app-functions}
Expert-designed analytic functions~\cite{wu2013towards, duggan2014contender, ahmad2011predicting, ahmad2009query} are used to predict concurrent runtime as a function of different resources (e.g., I/O, CPU, buffer pool status, and memory usage).
These methods generally consider a few types of resources, so we adapt and combine the merits of existing approaches to design an improved analytic function. 
This function considers a wide range of factors and contains parameters optimized by multi-dimensional regression.

\begin{table}[t]
	\centering
	\includegraphics[width=\columnwidth]{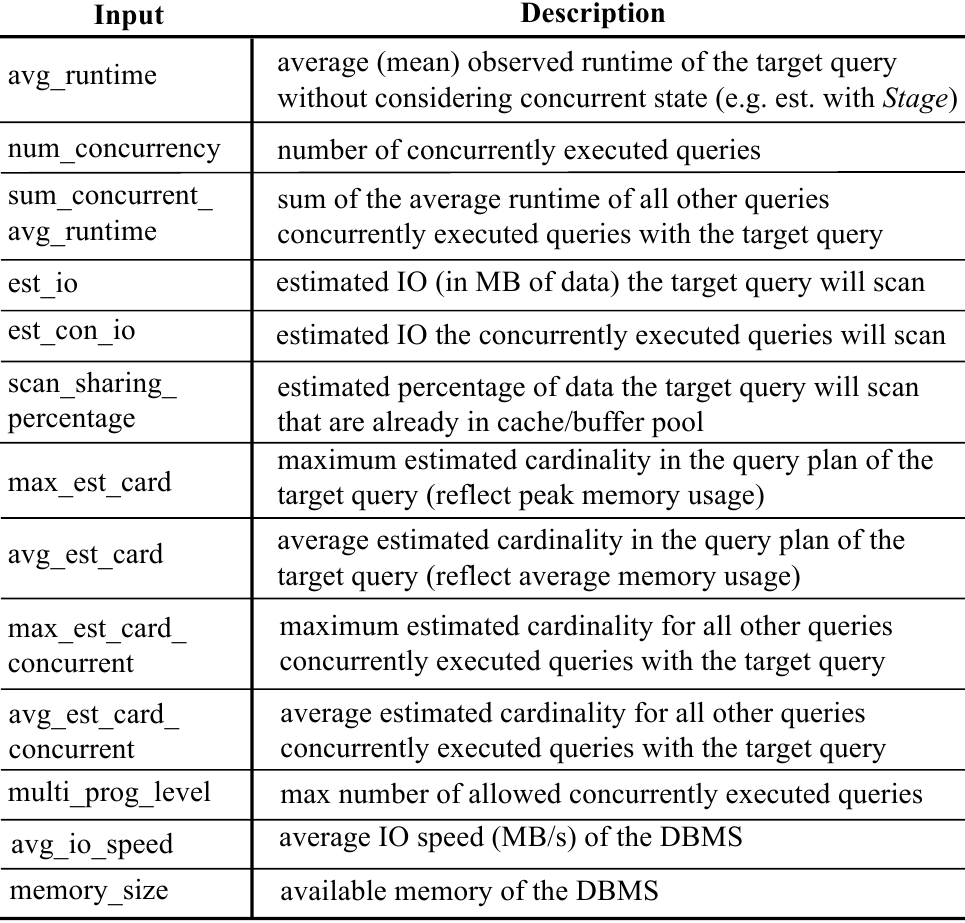}
	\caption{Input to analytical functions}
	\label{tab:input}
\end{table}

This improved analytic function considers a wide range of query features as its inputs, shown in Table~\ref{tab:input}. These features include the overall characteristics of the query, i.e., the average runtime of the target query, the number of concurrently executed queries, and the sum of the average runtime of all other queries concurrently executed queries with the target query.
We also provide characteristics related to the I/O usage of the queries, i.e., the estimated data access for the target query and its concurrently executed queries, the scan-sharing percentage.
To calculate the scan-sharing percentage, we estimate the cardinality of each table scan for the target query and its concurrently executed queries using the default DBMS cardinality estimation. 
This function also inputs characteristics related to the memory and CPU usage of the queries, i.e., maximum/average estimated cardinalities for the target query and its concurrently executed queries.
At last, we provide the function with system information, i.e., multi-programming level, average I/O speed, and memory size of the DBMS engine. 
It is worth noting that if the system information is not available, these system parameters can be set as optimizable parameters.

Then, this function takes these features as input to predict the runtime of the target query when concurrently executed with other queries. The details of this function can be found in Algorithm~\ref{alg:func}. 

\begin{algorithm}[h]
	\caption{Analytical function}
	\scriptsize
	\label{alg:func}
	\begin{algorithmic}[1]
		\Require function inputs are detailed in Table~\ref{tab:input}; optimizable parameters are specified as follows: \\
        {\color{blue} queueing\_weight}: adjust the queueing time of the target query \\
        {\color{blue} base\_io\_speed}: the base IO speed system allocates to each query \\
        {\color{blue} card\_adjustment}: adjust the cardinality estimation error\\
        {\color{blue} max\_mem\_weight}: adjust the effect of peak memory usage on the target query's runtime \\
        {\color{blue} avg\_mem\_weight}: adjust the effect of average memory usage on the target query's runtime \\
        {\color{blue} cpu\_usage}: adjust the effect of CPU usage on the target query's runtime
        \State num\_running\_queries = min(num\_concurrency + 1, multi\_prog\_level)
		\State running\_frac = num\_running\_queries/(num\_concurrency + 1)
        \LComment{fraction of running queries according to the MPL}
		\State queueing\_time = {\color{blue} queueing\_weight} * (1 - running\_frac) * sum\_concurrent\_avg\_runtime
        \LComment{estimated queueing time for target query due to MPL}
	  \State io\_speed = {\color{blue} 
         base\_io\_speed} + 
        avg\_io\_speed / num\_running\_queries
        \LComment{estimated IO speed of a query assuming each query has a base IO speed plus the IO speed due to contention, assuming that the system's average IO speed is evenly shared by all running queries}
        \State  io\_time = {\color{blue} card\_adjustment} * est\_scan * (1 - scan\_sharing\_percentage) / io\_speed
        \LComment{estimated time speed on IO accounting for data sharing}
        \State avg\_cpu\_time = avg\_runtime - io\_time
        \State cpu\_usage = (running\_frac * sum\_concurrent\_runtime) / avg\_runtime
	  \LComment{estimated proportion of CPU usage by the target query assuming the system allocates the CPU evenly to all running queries}
        \State max\_mem\_usage = max\_est\_card / (max\_est\_card\_concurrent  + max\_est\_card)
        \State avg\_mem\_usage = avg\_est\_card / (avg\_est\_card\_concurrent  + avg\_est\_card)
        \State mem\_usage = {\color{blue} 
         max\_mem\_weight} * max\_mem\_usage + {\color{blue} 
         avg\_mem\_weight} * avg\_mem\_usage 
        \LComment{estimated proportion of memory usage by the target query as a weighted sum of average memory usage and peak memory usage}
        \State cpu\_time = (1 + mem\_usage + {\color{blue} 
         cpu\_weight} * cpu\_usage) * avg\_cpu\_time
         \LComment{adjust the average CPU time of the target query by accounting for the resource usage from other concurrently running queries}
         \State runtime = queueing\_time + io\_time + cpu\_time
         \LComment{final runtime of a query is estimated to be the sum of queueing, IO, and CPU time}
        \State \Output runtime
	\end{algorithmic}
\end{algorithm}

This function separately estimates the queueing, IO, and CPU time of a target query, accounting for the effect of its concurrently running queries. 
Then, it estimates the system runtime of the target query as the sum of them.
It also contains several optimizable parameters to adjust for estimation errors and weight impact on different types of resources (i.e., IO, CUP, memory).
They are optimized jointly by Scipy package~\cite{virtanen2020scipy} using the training data.

\subsection{Additional examples of \sysname{}}
\label{sec:app-example}
\begin{figure}[h]
	\centering
	\includegraphics[width=8cm]{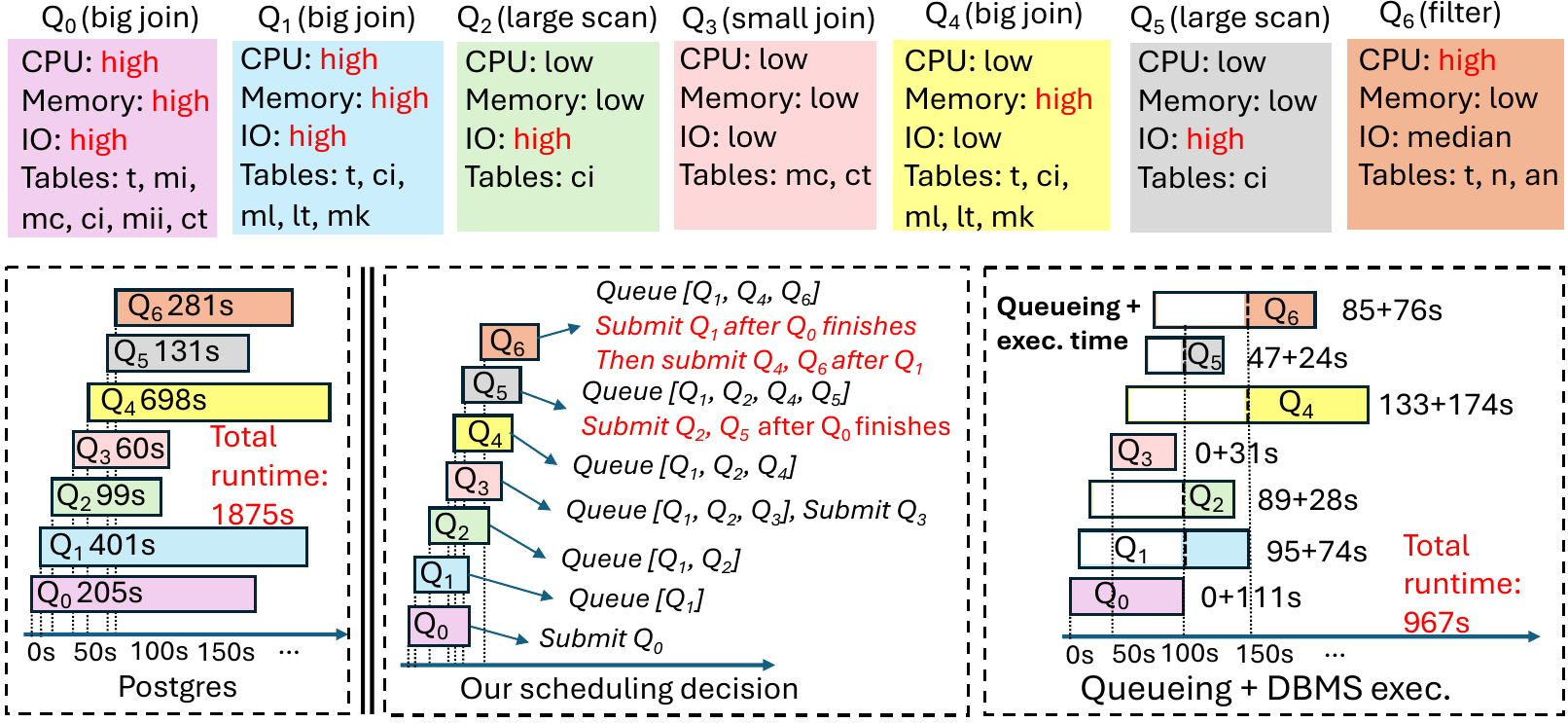}
	\caption{Example 3: deciding what is the optimal query to submit and when to submit it.}
	\label{fig:cs_optimal}
\end{figure}

\sysname{} may frequently add queries to the waiting queue after making the relevant predictions. 
Identifying the optimal queries (if any) to submit from the queued queries and when to submit them can significantly improve the overall query runtime.
Figure~\ref{fig:cs_optimal} shows a concrete example of seven queries $Q_0, \ldots, Q_6$ executed in our workloads. 
Brute-force executing them in Postgres will result in a heavy system load and long query runtime (1875s).
After making the relevant predictions,  \sysname{} believes it should add $Q_1, Q_2$ to the queue rather than submitting them.  When ingesting $Q_3$, $Q_1, Q_2$ understands submitting $Q_3$ now is the optimal decision because it has a minimal impact on the running query $Q_0$, and the current system state has a positive effect on $Q_3$ because of the shared table scan. 
Similarly, $Q_1, Q_2$ can decide the optimal queries and when to submit them for the rest of the queued queries, leading to a $2x$ speed-up over Postgres. 
These decisions are fundamentally based on \modelname{}'s fine-grained prediction of concurrently running queries, which the other baselines cannot do. 

\subsection{Adaptivity experiments}
\label{sec:app-adaptivity}

\noindent \textbf{Performance on BRAD queries executed in Redshift:}
\begin{figure}[h]
	\centering
	\includegraphics[width=8cm]{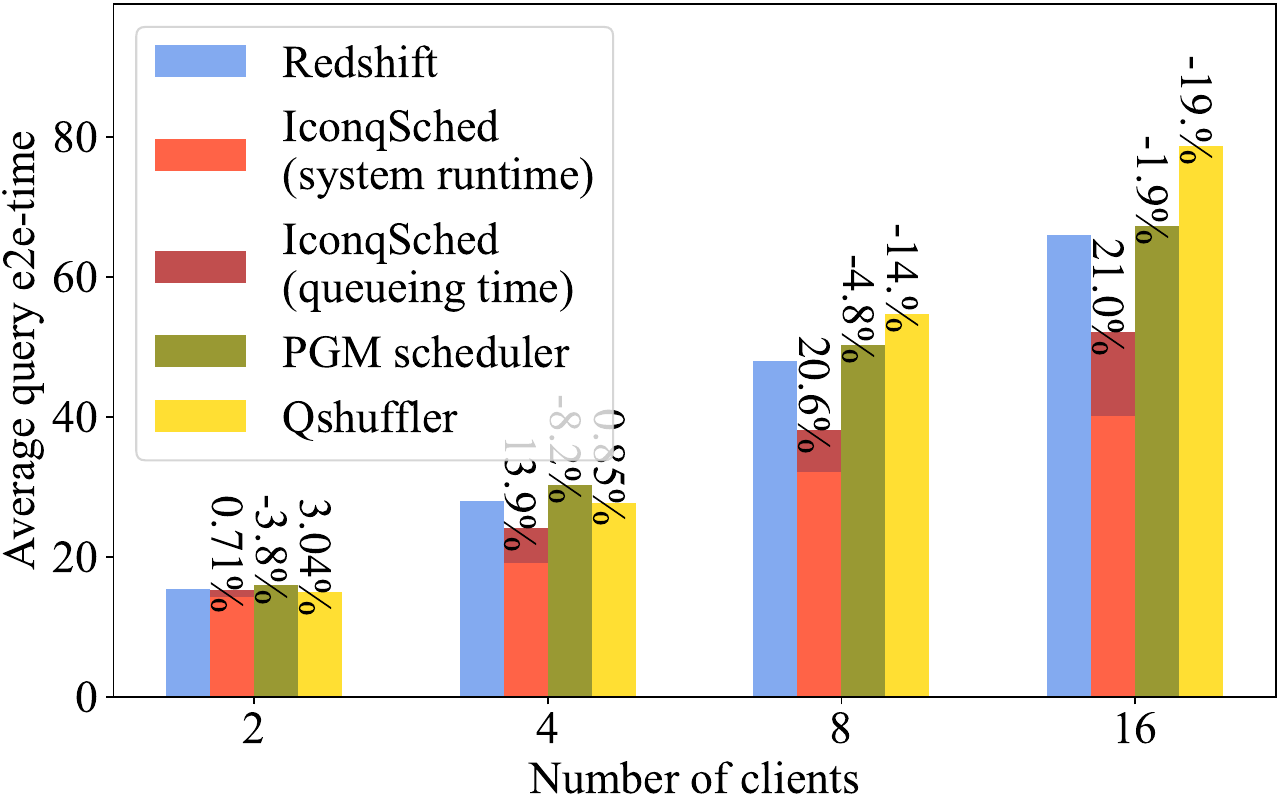}
	\caption{Performance of different schedulers with varied number of clients executing BRAD queries on Redshift.}
	\label{fig:adaptivity_redshift}
\end{figure}

Figure~\ref{fig:adaptivity_redshift} shows the average query runtime for different numbers of clients running BRAD queries on Redshift. For two clients, \sysname{} performs similarly to Redshift because there are not enough queries in the waiting queue for \sysname{} to select the query to submit that positively interacts with the running queries. 

\sysname{}'s performance improvement over Redshift increased from $2$ to $16$ clients because with more clients, more queries are waiting in the queue, and our scheduler has a better opportunity to submit the optimal queries at the optimal time. Moreover, with a median and heavy system load, cleverly managing the system resource usage can lead to significant improvement in execution time. 
For Qshuffler and PGM scheduler, it is difficult to understand the complex interaction of heavy system load using a heuristic-based runtime predictor or simple mechanism. 
Instead, they can sometimes increase the queueing time.
As a result, these two baselines frequently make suboptimal decisions and sometimes perform worse than Redshift.

\noindent \textbf{Performance on CAB queries:}
\begin{figure}[h]
	\centering
	\includegraphics[width=8cm]{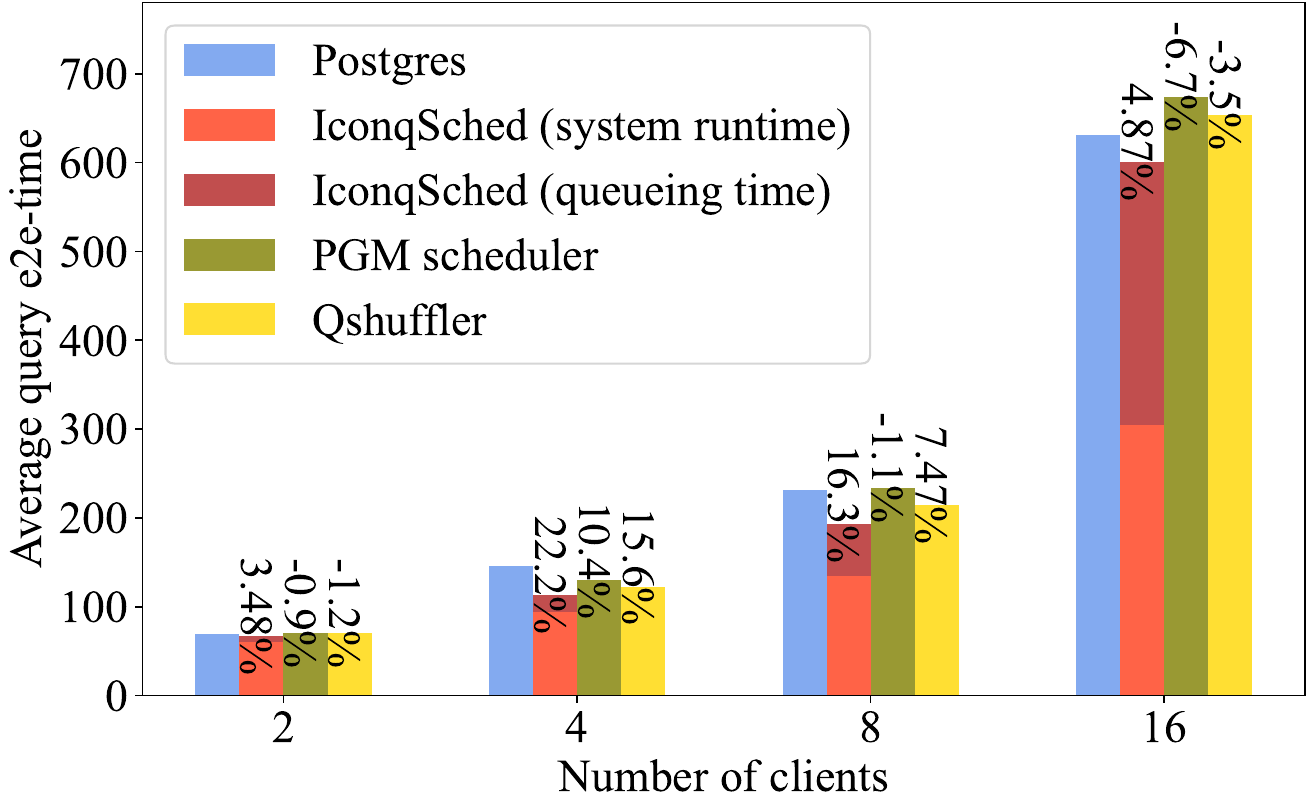}
	\caption{Performance of different schedulers with varied number of clients executing CAB queries on Postgres.}
	\label{fig:adaptivity_postgres_cab}
\end{figure}

\begin{figure}[h]
	\centering
	\includegraphics[width=8cm]{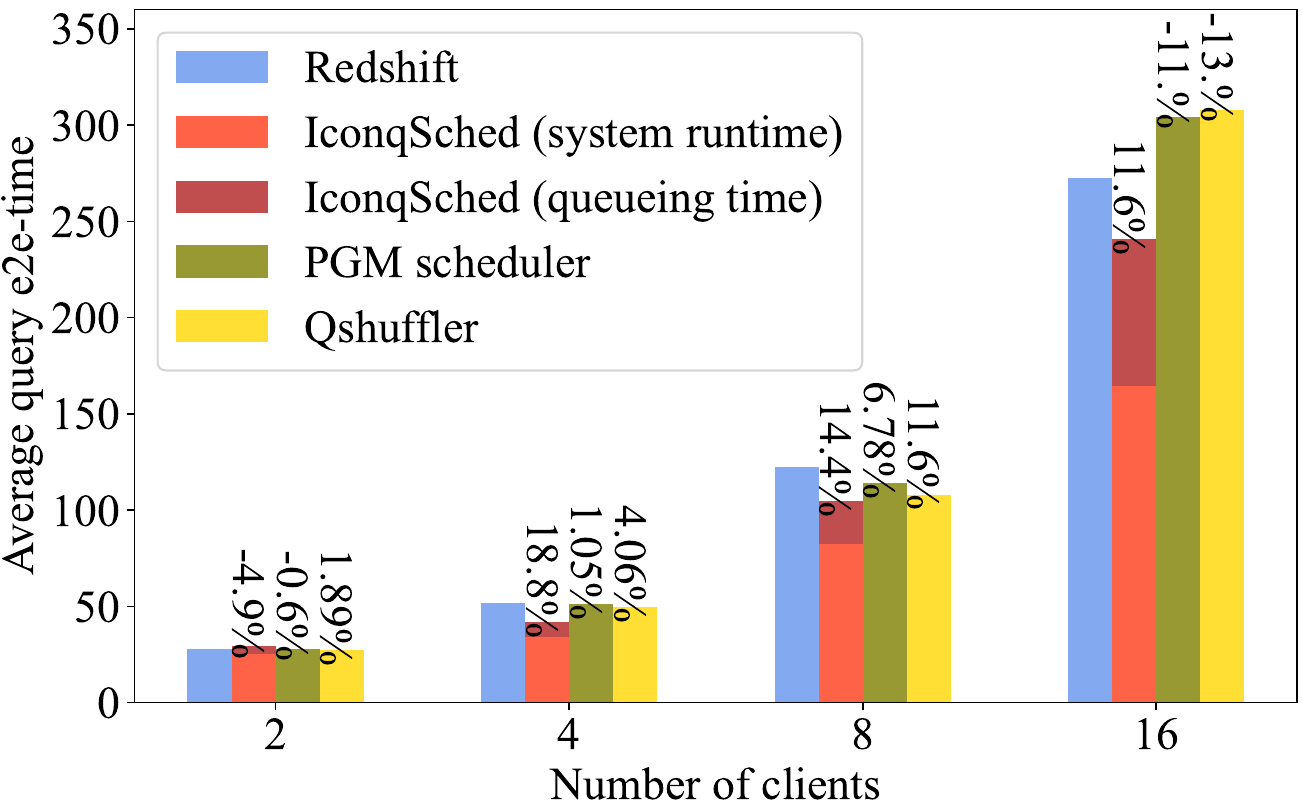}
	\caption{Performance of different schedulers with varied number of clients executing CAB queries on Redshift.}
	\label{fig:adaptivity_redshift_cab}
\end{figure}

Figure~\ref{fig:adaptivity_postgres_cab} and  Figure~\ref{fig:adaptivity_redshift_cab} shows the average query runtime for different numbers of clients running CAB queries in Postgres and Redshift, respectively. The observations are very similar to the performance of executing BRAD queries. 
It is worth noticing that $16$ clients issuing queries at the same time in Postgres severely overloads the system, resulting in an average query runtime of over $600s$ (last bar in Figure~\ref{fig:adaptivity_postgres_cab}).
In this case, \sysname{} tries to delay a large number of queries' execution to reduce the system load. 
However, since more queries are continuously arriving, many queued queries have to be submitted due to the starvation penalty (Equation 5). These queries are not necessarily the optimal queries to submit at the optimal time.
As a result, we see a significant improvement in the pure system execution time of these queries but not the end-to-end query time.

\subsection{Other aspects of runtime predictors}
\label{sec:app-runtime}

Table~\ref{tab:practicality} shows the training time, model size, and inference speed of all methods. 
We observe that the heuristic-based baselines (\textit{Stage}, \textit{Function}, and \textit{Qshuffler}) have much shorter training time, smaller model sizes, and faster inference speeds in comparison with the ML-based models (\modelname{} and \textit{GPredictor}). This additional overhead may be significant for short-running queries, but we believe it is affordable for heavy analytical workloads with many long-running queries. 
It is worth noticing that our method is only invoked for long-running queries as described in Section~\ref{sec:scheduling}. 

\begin{table}[h]
	\centering
	\caption{Other aspects of runtime predictors.}
	\scalebox{0.85}{
	\begin{tabularx}{\columnwidth}{Xlll}
		\toprule
		\textbf{Methods} & \textbf{Train time} & \textbf{Model size} & \textbf{Inference speed} \\
		& \textbf{(minutes)} & \textbf{(mb)} & \textbf{(ms/query)} \\
		\midrule
		\modelname{} & 148 & 3.8 & 17 \\
		Stage & 1 & 0.9 & 0.2 \\
		Function & 10 & 0.02 & 0.1 \\
		GPredictor  & 190 & 5.7 & 38 \\
		Qshuffler  & 1 & 0.8 & 0.2 \\
		\bottomrule
	\end{tabularx}
}
	\label{tab:practicality}
\end{table}